%% file: main_MEC_TCOM-arXiV.tex
\documentclass[journal,twoside]{IEEEtran}
\usepackage{graphicx}
\usepackage{cite,amssymb,amsmath,bm}
\usepackage{mathrsfs}
\usepackage[pagewise]{lineno}
\usepackage{color,soul}
\usepackage{tabulary}
\usepackage{multirow,multicol}
\usepackage{cases}
\usepackage{stfloats}
\usepackage{url}
\usepackage{booktabs}
\usepackage{algorithm,algpseudocode}
\usepackage{algorithmicx}
\usepackage[normalem]{ulem}
\usepackage{float}
\usepackage{tikz}
\usepackage{framed}
\usepackage{changepage}
\usepackage[bookmarks=false]{hyperref}
\ifCLASSOPTIONcompsoc
    \usepackage[caption=false, font=normalsize, labelfont=sf, textfont=sf, subrefformat=parens, labelformat=parens]{subfig}
\else
\usepackage[caption=false, font=footnotesize, subrefformat=parens, labelformat=parens]{subfig}
\fi

\algnewcommand{\IfThenElse}[3]{
  \State \algorithmicif\ #1\ \algorithmicthen\ #2\ \algorithmicelse\ #3}

\input{report-macros}

\def\BibTeX{{\rm B\kern-.05em{\sc i\kern-.025em b}\kern-.08em
    T\kern-.1667em\lower.7ex\hbox{E}\kern-.125emX}}

\algnewcommand{\IfThen}[2]{
  \State \algorithmicif\ #1\ \algorithmicthen\ #2}

\newcommand{\comment}[1]{}

\begin{document}
\begin{figure*}[t!]
\normalsize
This paper was submitted for publication in IEEE Transactions on Communications on January 22, 2023. It was finally accepted for publication on November 15, 2023.

\

\textcopyright~2023 IEEE. Personal use of this material is permitted. 
Permission from IEEE must be obtained for all other uses, in any current or future media, including reprinting/republishing this material for advertising or promotional purposes, creating new collective works, for resale or redistribution to servers or lists, or reuse of any copyrighted component of this work in other works.

\

Published in IEEE Transactions on Communications with open access and licensed under the Creative Commons Attribution 4.0 (CC BY 4.0). \textbf{Date of Publication:} 23 November 2023. \textbf{DOI:} 10.1109/TCOMM.2023.3336355.

\vspace{17cm}
\end{figure*}

\title{Joint Optimization of Uplink Power and Computational Resources in Mobile Edge Computing-Enabled Cell-Free Massive MIMO}
\author{Giovanni Interdonato and Stefano Buzzi
\thanks{This paper was supported by the Italian Ministry of Education University and Research (MIUR) Project ``Dipartimenti di Eccellenza 2018-2022'' and by the MIUR
PRIN 2017 Project ``LiquidEdge''. An excerpt of this article has been published in the proceedings of the 2022 IEEE International Conference on Communications (ICC)~\cite{Interdonato2021c}.
The authors are with the Department of Electrical and Information Engineering (DIEI) of the University of Cassino and Southern Lazio, 03043 Cassino, Italy (e-mail: giovanni.interdonato@unicas.it, buzzi@unicas.it), and with the Consorzio Nazionale Interuniversitario per le Telecomunicazioni (CNIT), 43124, Parma, Italy. S. Buzzi is also affiliated with Politecnico di Milano, Milano, Italy, and his work was also supported by the European Union under the \textit{Italian National Recovery and Resilience Plan} (NRRP) of NextGenerationEU, partnership on ``Telecommunications of the Future'' (PE00000001 - program ``RESTART'', Structural Project 6GWINET).}}

\maketitle

\begin{abstract}
The coupling of cell-free massive MIMO (CF-mMIMO) with Mobile Edge Computing (MEC) is investigated in this paper. A MEC-enabled CF-mMIMO architecture implementing a distributed user-centric approach both from the radio and the computational resource allocation perspective is proposed. A multi-objective optimization problem (MOOP) for the joint allocation of radio and remote computational resources is formulated, aimed at striking an optimal balance between total uplink power minimization and sum spectral efficiency maximization, under resource budget and latency constraints.  
In order to solve such a challenging non-convex problem, we convert the MOOP to an equivalent single-objective optimization problem (SOOP) through the weighted sum method and propose an iterative algorithm based on alternating optimization and sequential convex programming, along with an alternative heuristic resource allocation for distributed networks.
Finally,  we provide a detailed performance comparison between the proposed MEC-enabled CF-mMIMO architecture with its co-located counterpart, and its small-cell implementation. Numerical results reveal the effectiveness of the proposed resource allocation scheme, under different access point selection strategies, and the natural suitability of CF-mMIMO in supporting computation-offloading applications with benefits over users' transmit power and energy consumption, the effective latency experienced, and the computation offloading efficiency.
\end{abstract}

\section{Introduction}
\IEEEPARstart{T}{he recent} evolution of wireless networks has been characterized by an impressive growth not only of the amount of conveyed data traffic, but also of computationally-intensive applications with strict latency requirements for mobile devices. Applications such as online gaming, augmented reality and video image processing not only request extreme broadband connections, but also a considerable amount of computational power at the mobile devices.  A possible approach to indirectly increase the computing capabilities of the devices and prolong their battery lifetime is to (either fully or partially) delegate their computational tasks to the network, specifically to network entities known as \textit{network edge servers}\footnote{The edge servers are network entities figuratively placed at the \textit{edge} of the cellular access network, that is between the radio access network and the core network.}, in charge of collecting, processing and feeding data back to the users in a centralized fashion. This approach is known as \textit{mobile edge computing} (MEC) or mobile-edge computation offloading~\cite{Yang2008,Kumar2013,Sardellitti2015,Mach2017,Mao2017}. 

Cell-free massive multiple-input multiple-output (CF-mMIMO) is the ultimate embodiment of network MIMO~\cite{Ngo2017b,Interdonato2019,cellfreebook}.
CF-mMIMO is a technology based on the use of several distributed low-complexity access points (APs) that jointly serve the active users in their coverage area.  It inherits all the outstanding features of co-located massive MIMO \cite{redbook,massivemimobook}, such as nearly-optimal linear signal processing, predictable accurate performance, and simplified resource allocation and channel estimation, while providing additional key ingredients to theoretically achieve unprecedented levels of uniform data rates and ubiquitous connectivity: macro-diversity gain, inter-cell interference mitigation, and user proximity (see e.g.~\cite[Chapter 3.2]{interdonato2020cell} and references therein). Moreover, it is also amenable to scalable user-centric implementations~\cite{Buzzi2019c,Interdonato2019a,Bjornson2020}.

In this paper, we investigate the promising marriage between CF-mMIMO and MEC which share the same principle of bringing the resources (radio and computing, respectively) closer to the user. CF-mMIMO, thanks to its dense distributed topology and user-centric architecture, may greatly facilitate the computation offloading by enabling mobile devices to delegate either all or part of their computational tasks to multiple APs, each of which may be equipped with an edge server. 
Moreover, the central processing unit (CPU) of a CF-mMIMO system, which is generally equipped with a more powerful server, may serve as a backup edge computing to give, in turn, computation offloading support to the APs.
User proximity and the macro-diversity may significantly shorten the delay due to the computation offloading, thereby supporting stricter latency requirements, and reduce user's power consumption. Moreover, the user-centric approach ensures more uniform spectral efficiency (SE) and thereby the access to the remote computational resources may be indiscriminately granted to every user.
The ability of the network to accomplish users' computation offloading depends on how the radio and remote computational resources are allocated. This coupling calls for a joint optimization which is the main subject of this study.

\textbf{Related Works.} Many works on MEC optimize the interplay between the amount of computational tasks to offload, the latency due to the offloading process, and the energy consumption of mobile devices. First studies on MEC assumed simplified system models by considering either single-user systems~\cite{WZhang2013,YWang2016,Dinh2017} or interference-free multi-user systems~\cite{YMao2017,You2017}, focusing either on minimizing the energy consumption under latency constraints~\cite{You2017} or the delay due to the computation offloading under energy consumption constraints~\cite{Ren2018}.
Some of these works consider a \textit{binary computation offloading} model, wherein each device  executes its computational tasks either remotely or locally~\cite{WZhang2013,You2017}. Other studies assumed a more general \textit{partial computation offloading}~\cite{YWang2016,YMao2017,zhang2020communications} with only a fraction of computational tasks executed remotely.
An integrated framework for computation offloading and interference management in cellular networks was proposed in~\cite{CWang2017}.
All these works assumed single-antenna base stations (BSs).
 
More recently, MEC has been studied in conjunction with MIMO technologies. 
As an example, \cite{Sardellitti2015} considers a multi-cell MIMO system served by an edge server in a centralized fashion, and formulates a total energy minimization problem under latency and minimum rate constraints. 
Similarly, the MEC solution proposed in~\cite{QLi2018} focuses on minimizing the maximum latency of all the devices in a cloud-radio access network (C-RAN) with MIMO technology, while \cite{Nguyen2019} addresses the energy minimization problem accounting for imperfect channel state information (CSI) in a single-cell MIMO system. In~\cite{Sardellitti2018} an optimal association of mobile users to MEC resources is devised for a multi-user MIMO system with C-RAN architecture.
In~\cite{Pradhan2020} a successive inner convexification framework to minimize the total transmit power of the devices under latency constraints is proposed.
Several authors have also examined the coupling between MEC and massive MIMO.
The work~\cite{Hao2019} proposes a low-complexity algorithm to jointly optimize the radio and computing resources for a massive MIMO-enabled heterogeneous network with MEC. Similarly,~\cite{Zeng2020a, Zeng2020b} presented the effectiveness of employing massive MIMO for MEC, under zero-forcing combining, aiming at minimizing the maximum delay for offloading and
computing among the devices. The paper~\cite{YZhao2019}, instead, considers a massive MIMO system operating at the millimeter-wave (mmWave) frequency bands underlying traditional wireless local area networks with MEC. A dynamic computation offloading in MEC for ultra-reliable and low-latency communications at the mmWave frequency bands is proposed in~\cite{Merluzzi2020}.
The main common conclusion of these works is that multiple users can simultaneously offload their computational tasks by leveraging the additional degrees of freedom provided by massive MIMO, and the offloading efficiency, as well as the energy saving of the mobile devices, grows with the number of antennas of the massive MIMO BS.

Finally, the performance of CF-mMIMO with MEC has been recently explored in \cite{Femenias2022} and \cite{Mukherjee2020}. The former proposes a joint optimization of the partial offloading ratio per task and the resulting computational resources allocated at a single MEC server to minimize either the aggregated latency or the energy consumption at each user.
The author in~\cite{Mukherjee2020} investigate the successful edge
computing probability (SECP) for a target computation latency by using queuing theory and stochastic geometry, and by considering a random computation latency model.
The system model in~\cite{Mukherjee2020} consists of APs equipped with independent MEC servers, a CPU with a central MEC server (CS), and devices performing offloading either to the CS or to one of their serving APs, with some successful computing probability.
The joint decoding of the offloaded user data at one of the serving APs/CS is, however, not recommended due to extra delays caused by the fronthaul communications.
In fact, the considered architecture implements a specific instance of cell-free massive MIMO, namely a small-cells network.
Moreover, in~\cite{Mukherjee2020} both uplink transmit powers and allocated computational resources are fixed rather than optimized.
Following on this track, in this paper we explore the potential benefits of jointly optimizing radio and computational resources in a MEC-enabled CF-mMIMO system.
 
Our problem formulation is characterized by a multi-objective function and aims at striking an optimal balance between the minimization of the total uplink transmit power  and the maximization of the sum uplink SE.
Multi-objective optimization (MOO) is a mathematical framework to deal with optimization problems with multiple conflicting objective functions~\cite{Zadeh1963,Marler2004,Branke2008}. A survey of MOO applied to signal processing in wireless networks, with emphasis on massive MIMO systems and conflicting metrics such as SE, energy efficiency, coverage and total transmit power, was given in~\cite{Bjornson2014}. A conventional optimization approach consists in converting some of the objectives into constraints, whereas the fundamental approach of MOO consists in considering multiple objectives at once. There are two main methods: $(i)$ computing the sample points on the \textit{Pareto frontier} upon which making subjective decisions a posteriori, or $(ii)$ a priori converting the MOO problem (MOOP) to a single-objective optimization problem (SOOP) by combining the objectives into a suitable goal function (the most common is the \textit{weighted sum}) which reflects a subjective trade-off between metrics of interest. In the latter, the objectives are conventionally combined by using coefficients whose values reflect the subjective weight given to each metric. 

\textbf{Contributions:} Our technical contribution can be summarized as follows.
\begin{itemize}
\item We propose a MEC-enabled CF-mMIMO architecture implementing a user-centric approach both from the radio and the computational resource allocation perspective. Unlike prior studies investigating computation-offloading implementations in distributed networks~\cite{Hao2019,Pradhan2020,Mukherjee2020,Femenias2022}, our model considers that users' computational tasks can be divided into independent subtasks which can be remotely executed in a distributed fashion and in parallel at the MEC servers of properly selected APs and at the MEC server of the CPU.
\item We formulate an optimization problem for jointly allocating users' transmit powers and the remote computational resources for offloading. Unlike prior works~\cite{Hao2019,Zeng2020b,Pradhan2020,Mukherjee2020} we formulate a MOOP that optimizes the trade-off between total uplink transmit power minimization and sum SE maximization, under latency and resource budget constraints. 
\item For efficiently solving the non-convex MOOP, we formulate an equivalent SOOP by using the weighted sum method and devise a framework including alternating optimization and successive convex approximation (SCA) which, unlike prior works, accounts for: $(i)$ the user-centric cooperation clustering framework; $(ii)$ a distributed allocation of the computational resources; $(iii)$ a general formulation for any combining scheme and arbitrary correlated fading channels. As there is no unique solution for this SOOP, we provide its sub-optimal \textit{Pareto frontier}, namely the set of objectives corresponding to the Pareto sub-optimal solutions obtained by iteratively solving the SOOP for several values of the weights. Finally, we show how the weights in the SOOP indirectly determine the effective latency of the offloading process experienced by the users.
\item We propose an alternative low-complexity approach to the proposed joint resource allocation, which consists in heuristically allocating the MEC server computational resources to the users, and then optimizing with respect to the uplink powers.
\item Since the final solution achieved by SCA-based methods may depend on the feasible solution initialization, we present a method to properly initialize the proposed iterative optimization algorithm, and to provide a rigorous assessment on the problem feasibility.
\item For benchmarking purposes, we extend our joint optimization strategy to a multi-cell co-located massive MIMO system, and to a small-cell implementation of CF-mMIMO. The latter constitutes a deterministic variant of the framework described in~\cite{Mukherjee2020} wherein the task offloading model hinges on the knowledge of the users' computational demands and MEC servers' available computing resources. 
\item We provide a comprehensive simulation campaign to highlight the improvements introduced by the proposed MEC-enabled CF-mMIMO system in terms of: $(i)$ users' transmit power and energy consumption, $(ii)$ offloading latency, $(iii)$ amount of allocated remote computational resources, and $(iv)$ computation offloading efficiency. We also study its performance under different strategies of AP selection for providing the communication service.
\item A further insight about the effectiveness of the proposed joint uplink power and computational resource allocation (JPCA) scheme is provided by evaluating the interplay between energy consumption, allocated remote computational resources and offloading latency.   
\end{itemize}

\section{System Model} 
\label{sec:sysmodel}
We consider a CF-mMIMO system operating in time-division duplexing (TDD) mode and at sub-6 GHz frequency bands. A set of $L$ APs, equipped with $M$ antennas each, are geographically distributed and connected through a fronthaul network to a CPU. The APs coherently serve $K$ single-antenna users in the same time-frequency resources, with $LM\!\gg\!K$. 
The conventional block-fading channel model is considered, and let $\tau_c$ denote the channel coherence block length. In TDD mode, each coherence block accommodates uplink training, uplink and downlink data transmission, such that $\tauc\!=\!\tp\!+\!\tu\!+\!\td$, where $\tp$, $\tu$ and $\td$ are the training duration, the uplink and the downlink data transmission duration, respectively.  
 
Borrowing the notation of~\cite{Bjornson2020}, the channel between the $k$-th user and the $l$-th AP is denoted by the $M$-dimensional vector $\bh_{lk}$, with $\bh_{lk}\!\sim\!\CN(\bzero, \bR_{lk})$, and $\bR_{lk}\! \in\! \C^{M \times M}$ being the spatial correlation matrix. The corresponding large-scale fading coefficient is defined as $\beta_{lk}\! =\! \tr(\bR_{lk})/M$. 
The channel between the $k$-th user and all the APs in the system is obtained by stacking the channel vectors $\bh_{lk}$, $\forall l$ as $\bh_k\! =\! [\bh_{1k}\trans \cdots \bh_{Lk}\trans]\trans\! \in\! \C^{ML}$. The channel vectors of different APs are reasonably assumed to be independently distributed. As a consequence, we have $\bh_{k}\!\sim\!\CN(\bzero, \bR_{k})$, where $\bR_{k}\! =\! \text{blkdiag}(\bR_{1k}, \ldots, \bR_{Lk})\! \in\! \C^{ML \times ML}$ is user's $k$ block-diagonal spatial correlation matrix.   

\subsection{Centralized Uplink Training}
During the uplink training all the $K$ users synchronously send a pre-determined pilot sequence of $\tp$ samples. The pilot sequences are drawn by a set of $\tp$ orthonormal vectors. Specifically, $\sqrt{\tp} \bvphi_k\! \in\! \C^{\tp}$ denotes the pilot sent by the $k$-th user, with $\norm{\bvphi_k}\!=\!1$. Whenever $K$ is larger than $\tp$, the same pilot must be assigned to more than one user, causing pilot contamination. 
The pilot signal observed by AP $l$ is
$\bY_{\mathrm{p}, l} = \sum\nolimits^K_{j=1} \sqrt{\tp p_{\mathrm{p},j}}~\bh_{lj} \bvphi\trans_j + \bOmega_{\mathrm{p},l} \in \C^{M \times \tp},$ 
where $p_{\mathrm{p},j}$ is the transmit power of the uplink pilot symbol, and $\bOmega_{\mathrm{p},l}$ is a matrix of additive noise whose elements are independently distributed as $\CN(0,\sigma^2)$. For any user $k$, the $l$-th AP projects $\bY_{\mathrm{p}, l}$ along the $k$-th pilot sequence, which yields:
\begin{align}\label{eq:received_pilot_signal}
\!\!\by^{\text{p}}_{lk} &\!= \!\bY_{\mathrm{p},l}\bvphi^\ast_k \nonumber \\
&\!= \!\sqrt{\tp p_{\mathrm{p},k}}\bh_{lk} \!+\!\! \sum\limits^K_{j\neq k} \!\sqrt{\tp p_{\mathrm{p},j}}\bh_{lj} \bvphi\trans_j\bvphi^\ast_k \! + \!\bOmega_{\mathrm{p},l}\bvphi^\ast_k \!\in \!\C^{M},
\end{align}
where the second term captures the interference due to pilot contamination.
Assuming that the channel estimation is performed by the CPU, in each coherence interval, each AP needs to send the vector $\by^{\text{p}}_{lk}$ to the CPU.
Upon a prior knowledge of the channel correlation matrices, the CPU performs linear minimum-mean square error (MMSE) estimation of the $k$-th user channel $\bh_{lk}$ as
$\hat{\bh}_{lk} \!=\!  \sqrt{\tp p_{\mathrm{p},k}} \bR_{lk} \bPsi^{-1}_{lk} \by^{\text{p}}_{lk},$
where 
$\bPsi_{lk} \!=\! \EX{\by^{\text{p}}_{lk} (\by^{\text{p}}_{lk})\herm} \!=\! \tp \sum\nolimits_{j = 1}^K  p_{\mathrm{p},j} \bR_{lj}~|\bvphi_k\herm \bvphi_j|^2 \!+\! \sigma^2 \bI_M.$
The estimation error is independent of the estimate, and given by $\tilde{\bh}_{lk}\! =\! \bh_{lk}\! -\! \hat{\bh}_{lk}$. It is distributed as $\tilde{\bh}_{lk}\! \sim\! \CN(\bzero,\bC_{lk}),$ with $\bC_{lk}\! =\! \EX{\tilde{\bh}_{lk} \tilde{\bh}_{lk}\herm}\! =\! \bR_{lk}\! -\! \tp p_{\mathrm{p},k} \bR_{lk} \bPsi^{-1}_{lk} \bR_{lk}$.
Collecting all the channel estimates of user $k$ in a vector, we have $\hat{\bh}_k = [\hat{\bh}_{1k}\trans \cdots \hat{\bh}_{Lk}\trans]\trans$. Accordingly, it holds $\tilde{\bh}_{k}\! =\! \bh_{k}\! -\! \hat{\bh}_{k}\! \sim\! \CN(\bzero,\bC_k)$, with $\bC_k\! =\! \text{diag}(\bC_{1k}, \ldots, \bC_{Lk})$.

\subsection{User-Centric Uplink Data Transmission}
\label{subsec:uplink_data_transmission}
The uplink data signal received by AP $l$ is $ \by_l\! =\! \sum_{i=1}^K \bh_{li} s_i \!+\! \bn_l$, with $s_i$ being the data symbol transmitted by user $i$, $\EX{|s_i|^2}\! =\! p_i, \forall i$, and $\bn_l\! \sim\! \CN(0, \sigma^2 \bI_M)$ being the additive noise vector.
In a practical and scalable user-centric implementation each user is served by a subset of APs selected among those ensuring the best channel conditions. 
Let $\mathcal{M}_k$ be the set of the indices of the APs serving user $k$, and  
$\bD_{lk} \in \C^{M \times M}, \forall l, \forall k$ be a diagonal matrix such that $\bD_{lk}\! =\! \bI_M$, if $l\in\mathcal{M}_k$, $\bD_{lk}\! =\! \bzero_M$, otherwise.
Under centralized uplink operation, the CPU computes the estimate of the transmitted data symbol $s_k$ as
\begin{equation} \label{eq:signal-detection}
\hat{s}_k = \sum\limits_{l=1}^L \hat{s}_{lk} = \sum\limits_{l=1}^L \bv_{lk}\herm \bD_{lk} \by_l = \bv_{k}\herm \bD_{k} \by, 
\end{equation}
where $\bv_{lk}\! \in\! \C^M$ is the receive combining vector for the pair AP $l$-user $k$, $\bv_{k}\! = \! [\bv_{1k}\trans \cdots \bv_{Lk}\trans]\trans$, $\by\! =\! [\by_1\trans \cdots \by_L\trans]\trans\! \in\! \C^{ML}$, and $\bD_k\! =\! \text{diag}(\bD_{1k}, \ldots, \bD_{Lk})$. Eq. \eqref{eq:signal-detection} can be rewritten as
\begin{equation} \label{eq:signal-detection-2}
\hat{s}_k \!=\! \bv_{k}\herm \bD_{k} \hat{\bh}_k s_k \!+\! \bv_{k}\herm \bD_{k} \tilde{\bh}_k s_k \!+\! \sum\limits_{i \neq k}^K \bv_{k}\herm \bD_{k} \hat{\bh}_i s_i \!+\! \bv_{k}\herm \bD_{k} \bn,
\end{equation}
with $\bn \!=\! [\bn_1\trans \cdots \bn_L\trans]\trans \in \C^{ML}$ being the collective noise vector. The first term in~\eqref{eq:signal-detection-2} is the desired signal over the known partially estimated channel, the second term is the self-interference due to the (unknown) estimation error, the third term is the multi-user interference, and, lastly, the fourth term is the noise. An achievable uplink SE (bit/s/Hz) for user $k$, with centralized operation, is obtained by treating the last three terms as uncorrelated noise at the receiver:
\begin{align} 
\overline{\text{SE}}_k &\!=\! \frac{\tu}{\tc} \EX{\log_2 (1 + \text{SINR}_k)}, \quad \text{where} \label{eq:SE} \\
\text{SINR}_k &\!=\! \frac{p_k |\bv_{k}\herm \bD_{k} \hat{\bh}_k|^2}{\sum\nolimits_{i \neq k}^K p_i |\bv_{k}\herm \bD_{k} \hat{\bh}_i|^2 \!+\! \bv_{k}\herm \bZ_k \bv_{k} \!+\! \sigma^2 \norm{\bD_k\bv_{k}}^2} \label{eq:SINR} \, ,
\end{align}
with $ \bZ_k = \sum\nolimits_{i=1}^K p_i  \bD_k \bC_i \bD_k.$
This achievable SE holds for any combining scheme and arbitrary correlated fading channels, and accounts for user-centric data detection, channel estimation error, pilot contamination and estimation overhead. 
Hereafter, we consider the so-called Partial MMSE (P-MMSE) combining~\cite{Bjornson2020} which guarantees scalability and an excellent trade-off between performance and computational complexity. For an arbitrary user $k$, P-MMSE suppresses only the strongest interference contributions which are caused by the users whose indices are in the set $\mathcal{S}_k = \{i: \bD_k \bD_i \neq \bzero_{LM} \}$. The P-MMSE collective combining vector is given by    
\begin{equation} \label{eq:PMMSE-combining-vector}
\bv_k^{\mathsf{P-MMSE}} \!=\! p_k \Bigg(\sum\limits_{i \in \mathcal{S}_k} p_i \bD_k \hat{\bh}_i \hat{\bh}_i\herm \bD_k \!+\! \bZ_{\mathcal{S}_k} \!+\! \sigma^2 \bI_{LM} \!\Bigg)^{\!\!\!-1} \!\bD_k \hat{\bh}_k \, ,
\end{equation}
where $\bZ_{\mathcal{S}_k} \!=\! \sum\nolimits_{i \in \mathcal{S}_k} p_i  \bD_k \bC_i \bD_k$.

\section{Computation-Offloading and Latency Model}
\label{sec:offloading-model}
We assume that both the APs and the CPU\footnote{The CPU is either a physical or a logical entity, an edge-cloud processor located in the same geographical area as the APs.} offer computational facility to the users.
Each user has a set of computational tasks to offload to multiple distributed MEC servers on some  APs and/or to the MEC server at the CPU. In particular, we denote by $\mathcal{G}_k$ the set of MEC servers at the APs and CPU that can provide computational offloading service to user $k$. Let $\mathcal{O}_k \subseteq \mathcal{G}_k$ be the set of APs where user $k$'s computational tasks can be offloaded\footnote{Notice that there is no implicit relation between $\mathcal{O}_k$ and 
$\mathcal{M}_k$, even though it is reasonable to expect that $\mathcal{M}_k \subseteq 
\mathcal{O}_k$.}. In a small system with one CPU, as it is the case for this paper, it is also reasonable that $\mathcal{O}_k$ coincides with the set of all the APs in the system.
We assume that user $k$ needs to execute one or more computational tasks within a maximum tolerable latency $\mathcal{L}_k$. All the relevant information on these computational tasks can be encoded into $b_k$ bits, and their execution requires a total of $w_k$ computation cycles, that can be decomposed in $T_k$ computational subtasks, consisting of $w_{k,1}, w_{k,2}, \ldots, w_{k,T_k}$ computation cycles, with $\sum_{t=1}^{T_k} w_{k,t}=w_k$.
AP $l$ has a computational capability of $f^{\mathsf{AP}}_l$ computation cycles per second (\textit{computational rate}). While, the CPU can execute up to $f^{\mathsf{CPU}}$ computation cycles per second. The fractions of computational resources assigned to subtask $i$ of user $k$ by the generic AP $l$ and by the CPU are denoted by $f^{\mathsf{AP}}_{l,k}(i)$ and $f^{\mathsf{CPU}}_k(i)$, respectively\footnote{The optimization problem that will be considered in this paper will enforce the constraint that each subtask is entirely executed at only one AP or at the CPU, i.e., subtasks cannot be further divided into parallel sub-subtasks to be executed in different MEC servers.}.
\begin{figure}[!t]
\centering
\resizebox{.98\columnwidth}{!}{%
\input{signalling-diagram}%
}
\vspace{-3mm}
\caption{Signalling diagram describing the computational offloading process for an arbitrary user $k$. The computational offloading may occur at the MEC servers of both the CPU and the APs.\vspace{-5mm}}
\label{fig:signalling-diagram}
\end{figure}
Accordingly, it holds
\begin{align*}
\sum\limits_{k=1}^K \sum\limits_{i=1}^{T_k}  f^{\mathsf{CPU}}_k(i) \leq f^{\mathsf{CPU}},~\text{and}~\sum\limits_{k=1}^K\sum\limits_{i=1}^{T_k} f^{\mathsf{AP}}_{l,k}(i) \leq f^{\mathsf{AP}}_l,~ \forall l.
\end{align*} 
The total amount of remote computational resources assigned to subtask $i$ of user $k$ is given by $f_k(i) \!=\! f^{\mathsf{CPU}}_k(i) \!+\! \sum\limits_{l \in \mathcal{O}_k} f^{\mathsf{AP}}_{l,k}(i)\,.$ Then, $w_{k,i}/f_k(i)$ represents the computational time needed to execute $w_{k,i}$ cycles  (\textit{computational latency}). 
Since the $T_k$ computational subtasks for user $k$ are executed in parallel, the resulting computational latency for the task of user $k$ is $\max\limits_{i=1, \ldots, T_k} {w_{k,i}}/{f_k(i)}\,$.

While, the amount $b_k/R_k$ is the time needed to transmit $b_k$ bits to the APs (\textit{transmission latency}) over the wireless channel  supporting a rate $R_k \!=\! B  \times  \text{SE}_k$, with $B$ being the transmission bandwidth and $\text{SE}_k$ being the instantaneous SE, i.e., the value attained by~\eqref{eq:SE} with no expectation.
Lastly, an additional latency contribution (\textit{fronthaul latency}) is due to the forwarding of the $b_k$ bits from all the APs in the set $\mathcal{M}_k$ to the CPU over the fronthaul network, which, assuming synchronous transmission across the APs, amounts to $2 b_k M \xi/C_\mathsf{FH}$, where $\xi$ denotes the number of bits used to quantize both real and imaginary parts of the uplink data signal $\by$, and $C_\mathsf{FH}$ is the fronthaul capacity of the link between any AP and the CPU, expressed in bit/s.    
Hence, the computational offloading must fulfill the following latency constraint~\cite{Pradhan2020}
\begin{equation} \label{eq:latency-constraint}
\frac{b_k}{R_k} + \displaystyle \max_{i=1, \ldots, T_k} \frac{w_{k,i}}{f_k(i)} + \frac{2 b_k M \xi}{C_\mathsf{FH}} \leq \mathcal{L}_k, \forall k,
\end{equation}
where we assume that $\mathcal{L}_k$ includes any delay related to the signalling between AP and CPU, and the time needed to send the computational output back to the user. 
The latency constraint in~\eqref{eq:latency-constraint} clearly couples radio and computational resources.
\Figref{fig:signalling-diagram} illustrates the signalling diagram of the computational offloading process for an arbitrary user $k$. As a first step, user $k$ sends on the air interface a \textit{computational offloading request} followed by the symbols encoding the data to be processed, the program to be executed remotely, and, also, the details on the subtasks which the main computational task is composed of. It is assumed that the overall computational offloading message has a length of $b_k$ bits.
These $b_k$ bits are treated as normal information data, so
they are sent from the user $k$ during the uplink data transmission phase; the corresponding received signals are locally processed at the APs serving user $k$ (i.e., the APs in the set $\mathcal{M}_k$), and sent to the CPU over the fronthaul network for the centralized receive combining and the data decoding. The CPU,
based on the received computational requests from all the users in the network, and on the knowledge of the estimated uplink channels, and of the available computing power at the CPU itself and at the APs, allocates each \textit{batch job}, represented by an entry of the set $\left\{ w_{k,i} \, : \, k=1, \ldots, K; i=1, \ldots, T_k \right\}$,  either to itself or to one AP.
The allocation of the computational resources between CPU and APs is subject to optimization as detailed in the next subsection. Once received the computational output of each batch job from the APs, the CPU sends the combined computational output back to the APs in the set $\mathcal{M}_k$, which finally perform a joint downlink data transmission to user $k$.

\subsection{\!Joint Power and Computational Resource Allocation\! (JPCA)\!}
\label{subsec:JPCA}
We jointly allocate the users' transmit powers and the remote computational resources assigned to the users aiming at simultaneously minimizing the total uplink transmit power and maximizing the sum SE. These are two conflicting objective functions of a MOOP which will be treated through the classical scalarization technique. In particular, the MOOP is converted into a SOOP designing a single goal function that reflects a pre-determined trade-off between the objectives. 
To this end, let us first introduce the vector of the uplink powers $\bp = [p_1 \cdots p_K]\trans$ and
the set $\mathcal{F}$ containing the allocated computational resources, that is
\begin{align} \label{eq:set-F}
   \mathcal{F} &= \{ f^{\mathsf{CPU}}_k(i),\ f^{\mathsf{AP}}_{l,k}(i): \, l=1, \ldots, L; \, k=1, \ldots, K; \nonumber \\
   &\qquad i=1, \ldots, T_k \}\, .
\end{align}
The SOOP for the proposed JPCA can be formulated as
\begin{subequations} \label{prob:P1}
\begin{align}	
  \mathop {\text{min}}\limits_{\substack{\bp,~\mathcal{F} \\ \bnu}} & \; \vpp \boldsymbol{1}\trans_K \bp - \vpse\boldsymbol{1}\trans_K \bnu  \label{prob:P1:obj} \\[-1.5ex]
  \text{s.t.} &\; \frac{b_k}{B~\text{SE}_k(\bp)} \!+\! 
  \max_{j=1, \ldots, T_k} \frac{w_{k,j}}{f^{\mathsf{CPU}}_k(j) \!+\!\! \sum\limits_{l \in \mathcal{O}_k} f^{\mathsf{AP}}_{l,k}(j)}\leq \mathcal{\widetilde{L}}_k, \forall k, \label{prob:P1:C1} \\[-1ex] 
  			  &\; \text{SE}_k(\bp) \geq \nu_k,~\forall k,\label{prob:P1:C2} \\[-.5ex]
  			  &\; \sum\nolimits_{k=1}^K \sum\nolimits_{i=1}^{T_k} f^{\mathsf{CPU}}_k(i) \leq f^{\mathsf{CPU}}, \label{prob:P1:C3} \\[-.5ex]
  			  &\; \sum\nolimits_{k=1}^K \sum\nolimits_{i=1}^{T_k} f^{\mathsf{AP}}_{l,k}(i)\leq f^{\mathsf{AP}}_l, \; \forall l, \label{prob:P1:C4} \\[-.5ex]
  			  &\; u(f^{\mathsf{CPU}}_k(i)) + \sum\nolimits_{l \in \mathcal{O}_k} u(f^{\mathsf{AP}}_{l,k}(i))=1, \, \forall k, \forall i, \label{prob:P1:C5} \\[-.5ex]
   			  &\; \bzero_K \preceq \bp \preceq p_{\text{max}}\cdot \boldsymbol{1}_K, \label{prob:P1:C6} \\[-.5ex]
  			  &\; f \in \mathbb{R}_{\geq0}, \; \forall f \in \mathcal{F}\, . \label{prob:P1:C7} 
\end{align}
\end{subequations}
In the above problem,  $\mathcal{\widetilde{L}}_k \!=\! \mathcal{L}_k \!-\! (2 b_k M \xi / C_\mathsf{FH})$, $p_{\text{max}}$ is the maximum transmit power per user, $\nu_k$ represents the minimum instantaneous SE target for user $k$ and $\bnu \!=\! [\nu_1 \cdots \nu_K]\trans$. Constraints \eqref{prob:P1:C3} and \eqref{prob:P1:C4} ensure that the allocate computational cycles do not exceed the computational capacity of the CPU and the APs, respectively, while  constraint \eqref{prob:P1:C5}, where $u(\cdot)$ denotes the unit-step function, ensures that each computational subtask is executed in one processing unit only, i.e. either at the CPU or in one of the APs. Constraint \eqref{prob:P1:C7} ensures that the allocated cycles are positive real numbers, and possible non-integer optimal solutions are then rounded (i.e., continuous relaxation).
Moreover,
\begin{align}\label{eq:weights}
\vpp = \frac{\op}{Kp_{\text{max}}}, \quad \vpse = \frac{\ose}{K \max\nolimits_k \text{SE}^{(0)}_k},
\end{align}
where $\{\text{SE}^{(0)}_k\}$ are constants denoting reference instantaneous SEs attained by a pre-determined uplink power allocation (e.g., setting $p_k \!=\! p_{\text{max}}, \; \forall k$) and $\op,~\ose \in [0,1]$ so that each term of the objective function~\eqref{prob:P1:obj} is dimensionless and takes on values in the interval $[0,1]$.
These weights determine a trade-off between the minimization of the total trasmit power and the maximization of the sum SE, hence they indirectly act over the effective offloading latency, which is given by the l.h.s. of the constraint~\eqref{prob:P1:C1}.
Problem~\eqref{prob:P1} is clearly non-convex with respect to $\bp$ due to the non-convexity (non-concavity) of the latency constraint~\eqref{prob:P1:C1} and the minimum SE constraint~\eqref{prob:P1:C2}. Moreover, the non-linear constraint \eqref{prob:P1:C5} makes the optimization problem a mixed integer one.
To solve the above problem, we resort to the alternating optimization approach, i.e. we first set some initialization values for $\bp$ and $\bnu$ and solve the problem with respect to $\mathcal{F}$, and then for the obtained value of $\mathcal{F}$, we solve the problem with respect to $\bp$ and $\bnu$. The process is iterated until the value attained by the objective function converges and/or a maximum number of iteration has been reached. Notice that, since at each step the value of the optimized variable is updated only if it leads to a smaller value of the objective function, and since the objective function is bounded from below, the procedure provably converges.

\subsubsection{Optimization with respect to $\mathcal{F}$}
We focus on the problem of determining the frequencies in $\mathcal{F}$ for fixed values of $\bp$ and $\bnu$. First of all, notice that the objective function in \eqref{prob:P1} does not depend on $\mathcal{F}$. The purpose of the optimization with respect to the computational cycles is thus to make the second term in \eqref{prob:P1:C1} as small as possible so that to make the latency constraint as loose as possible, and the subsequent optimization with respect to $\bp$ and $\bnu$ can be done on a wider search domain. Based on the above reasoning, we consider the following problem:
\begin{subequations} \label{prob:P1_F}
	\begin{align}	
		\mathop{\text{minimize}}\limits_{k=1, \ldots, K,~\mathcal{F}} & \quad \displaystyle \max_{j=1, \ldots, T_k} \frac{w_{k,j}}{f^{\mathsf{CPU}}_k(j) + \sum_{l \in \mathcal{O}_k} f^{\mathsf{AP}}_{l,k}(j)}  \label{prob:P1_F:obj} \\[-0.2ex]
		\text{s.t.} &\quad  {f^{\mathsf{CPU}}_k(j) + \sum_{l \in \mathcal{O}_k} f^{\mathsf{AP}}_{l,k}(j)} \geq \frac{w_{k,j}}{\mathcal{\widetilde{L}}_k-\frac{b_k}{B~\text{SE}_k(\bp)}}, \nonumber \\
		&\qquad\forall k, \, \forall j=1, \ldots, T_k, \label{prob:P1_F:C1}\\[-.5ex]
		&\quad \eqref{prob:P1:C3}, \eqref{prob:P1:C4}, \eqref{prob:P1:C5}, \eqref{prob:P1:C7} \; .
	\end{align}
\end{subequations}
In the above formulation, constraint \eqref{prob:P1_F:C1} descends from, and is equivalent to, constraint \eqref{prob:P1:C1}. An equivalent formulation for problem \eqref{prob:P1_F} is the following:
\begin{subequations} \label{prob:P1_F2}
	\begin{align}	
		\mathop {\text{minimize}}\limits_{\mathcal{F}} & \quad t  \\[-1ex]
		\text{s.t.} &\quad  
		{f^{\mathsf{CPU}}_k(j) + \sum_{l \in \mathcal{O}_k} f^{\mathsf{AP}}_{l,k}(j)} \geq \frac{w_{k,j}}{\mathcal{\widetilde{L}}_k-\frac{b_k}{B~\text{SE}_k(\bp)}}, \nonumber \\
		&\qquad \forall k, \, \forall j=1, \ldots, T_k, \label{prob:P1_F2:C1} \\[-.2ex] 
		&\quad {f^{\mathsf{CPU}}_k(j) + \sum\nolimits_{l \in \mathcal{O}_k} f^{\mathsf{AP}}_{l,k}(j)} \geq \displaystyle \frac{w_{k,j}}{t}, \nonumber \\
		&\qquad \forall k, \, \forall j=1, \ldots, T_k, \label{prob:P1_F2:C2} \\[-.5ex] 
		&\quad \eqref{prob:P1:C3}, \eqref{prob:P1:C4}, \eqref{prob:P1:C5}, \eqref{prob:P1:C7} \; . 
	\end{align}
\end{subequations}
Problem \eqref{prob:P1_F2} is a feasibility program; specifically, the goal is to find the optimal frequencies in $\mathcal{F}$ that minimize the value of $t$ and such that the problem \eqref{prob:P1_F2} admits a non-empty feasible set. This can be accomplished through the \textit{bisection method}~\cite{Boyd2004}. In particular, feasibility 
should be first checked assuming that $t$ is unboundedly large, i.e., which makes constraints \eqref{prob:P1_F2:C1} inactive. If the problem reveals to be feasible with $t \rightarrow +\infty$, then the bisection method can be applied; specifically,  
elaborating on the constraints 
\eqref{prob:P1_F2:C1} and \eqref{prob:P1_F2:C2}, the following values can be used as the start and the end of the initial search interval on $t$:
\begin{align}
\begin{split}
	t_0 &= 0.9 \frac{\max_{k,j} w_{k,j}}{\max\{f^{\mathsf{CPU}}, f^{\mathsf{AP}}_1, \ldots, f^{\mathsf{AP}}_L  \}} \, , \\
	t_1 &= 1.1 \max_{k} \left[ \mathcal{\widetilde{L}}_k-\frac{b_k}{B~\text{SE}_k(\bp)} \right].	
\end{split}
\label{eq:bisection_interval}
\end{align} 
Notice that without constraint \eqref{prob:P1:C5}, \eqref{prob:P1_F2} would be a simple linear program which could be solved with any off-the-shelf optimization routine. The presence of \eqref{prob:P1:C5}, instead, requires further efforts. For a preassigned value of $t$, namely for each iteration of the bisection algorithm, we consider the following associated problem to ascertain the feasibility of the constraints in \eqref{prob:P1_F2}:
\begin{subequations} \label{prob:P1_F3}
	\begin{align}	
		\mathop {\text{max}}\limits_{\mathcal{F}} & \; \displaystyle
		\sum_{k=1}^K \sum_{j=1}^{T_k} \left(f^{\mathsf{CPU}}_k(j) +\sum_{l \in \mathcal{O}_k} f^{\mathsf{AP}}_{l,k}(j)\right) \label{prob:P1_F3:obj}  \\[-0.2ex]
		\text{s.t.} &\;  
		{f^{\mathsf{CPU}}_k(j) \!+\!\! \sum_{l \in \mathcal{O}_k} f^{\mathsf{AP}}_{l,k}(j)} \!\geq\! \max\left\{\frac{w_{k,j}}{t}, \frac{w_{k,j}}{\mathcal{\widetilde{L}}_k \!-\!\frac{b_k}{B~\text{SE}_k(\bp)}}\right\}, \nonumber \\
		&\quad\;\forall k, \, \forall j=1, \ldots, T_k, \label{prob:P1_F3:C1} \\[-.5ex] 
		&\; \eqref{prob:P1:C3}, \eqref{prob:P1:C4}, \eqref{prob:P1:C5}, \eqref{prob:P1:C7} \; . 
	\end{align}
\end{subequations}
Problem \eqref{prob:P1_F3} aims at finding both the optimal user-to-MEC-server association for the computation offloading and the optimal amount of remote computational resources to be assigned. This problem can be shown to be cast as a \textit{multiple knapsack problem}. Firstly, we introduce $\dot{f}_{\ell,k}(j),~\ell \!\in\! \mathcal{G}_k$, to denote the computational resources that can be allocated to user $k$ for offloading its subtask $j$ on MEC server $\ell$. Hence, $\dot{f}_{\ell,k}(j)$ is equal to either $f^{\mathsf{CPU}}_k(j)$ or $f^{\mathsf{AP}}_{l,k}(j)$, with $l \!\in\! \mathcal{O}_k$. Then, we introduce
$\mathcal{X} \!=\!\left\{ x_{\ell,k,j} \!\in\! \{0,1\}: \, k=1, \ldots, K; \, j=1, \ldots, T_k; \,  \ell \in \mathcal{G}_k  \right\}$ to denote the set of the binary variates $\{x_{\ell,k,j}\}$ mathematically handling the user-to-MEC-server association. Hence, the amount of computational resources effectively allocated to user $k$ for its subtask $j$ is given by $\sum\nolimits_{\ell \in \mathcal{G}_k} \dot{f}_{\ell,k}(j) x_{\ell,k,j}$, with $x_{\ell,k,j} \!\in\! \{0,\,1\}$ and $\sum\nolimits_{\ell \in \mathcal{G}_k} x_{\ell,k,j} \!=\! 1$, since each subtask is executed either at the CPU or in one of the APs. Therefore, problem \eqref{prob:P1_F3} can be rewritten as
\begin{subequations} \label{prob:P1_F4}
\begin{align}	
		\mathop {\text{max}}\limits_{\mathcal{F},~\mathcal{X}} & \;
		\sum\limits_{k=1}^K \sum\limits_{j=1}^{T_k} \sum\limits_{\ell \in \mathcal{G}_k} \dot{f}_{\ell,k}(j) x_{\ell,k,j}  \label{prob:P1_F4:obj}  \\[-0.2ex]
		\text{s.t.} &\;  
		\sum\limits_{\ell \in \mathcal{G}_k} \dot{f}_{\ell,k}(j) x_{\ell,k,j} \!\geq\! 
		\max\left\{\frac{w_{k,j}}{t},  \frac{w_{k,j}}{\mathcal{\widetilde{L}}_k \!-\!\frac{b_k}{B~\text{SE}_k(\bp)}}\right\}, \nonumber \\
		&\quad\; \forall k, \, \forall j=1, \ldots, T_k,\label{prob:P1_F4:C1} \\[-.5ex] 
		&\; \eqref{prob:P1:C3}, \eqref{prob:P1:C4}, \eqref{prob:P1:C5}, \eqref{prob:P1:C7} \; , 
\end{align}
\end{subequations}
and $\mathcal{F}$ is redefined as $$\mathcal{F} \!=\! \left\{ \dot{f}_{\ell,k}(j)\,: \, k\!=\!1, \ldots, K; \, \ell\in\mathcal{G}_k; \, j\!=\!1, \ldots, T_k \right\}\, .$$
Secondly, we avoid optimizing the amount of computational resources in \eqref{prob:P1_F4} by setting $\dot{f}_{\ell,k}(j)$ to a constant as 
$$ \xi_{k,j}(t, \bp) \!\triangleq\!  \max\left\{\dfrac{w_{k,j}}{t},  \dfrac{w_{k,j}}{\mathcal{\widetilde{L}}_k-\frac{b_k}{B~\text{SE}_k(\bp)}}\right\}\,, \forall\,\ell \!\in\! \mathcal{G}_k.$$ 
By doing so, constraint~\eqref{prob:P1_F4:C1} becomes inactive, and the amount of computational resources to assign to each user will be eventually determined at the end of the bisection algorithm, being possibly dependent on the smallest value of $t$. While, for each iteration of the bisection algorithm, we only optimize the user-to-MEC-server offloading association. Hence, problem \eqref{prob:P1_F4} can be reformulated as the following multiple knapsack problem:
\begin{subequations} \label{prob:knap}
	\begin{align}	
		\mathop {\text{maximize}}\limits_{\mathcal{X}} & \quad \sum\limits_{k=1}^K \sum\limits_{j=1}^{T_k}\sum\limits_{\ell\in \mathcal{G}_k} \xi_{k,j}(t,\bp) x_{\ell,k,j} \label{prob:knap:obj}\\[-0.2ex]
		\text{s.t.} &\quad  
		\sum\limits_{k=1}^K \sum\limits_{j=1}^{T_k} \xi_{k,j}(t, \bp) x_{\ell,k,j} \leq \dot{f}_{\ell} \, , \quad \forall \ell, \label{prob:knap:C1}\\[-0.5ex] 
		&\quad \sum\limits_{\ell \in \mathcal{G}_k} x_{\ell,k,j} \leq 1 \, , \quad \forall k, \forall j= 1, \ldots, T_k \label{prob:knap:C2} \; , \\[-0.2ex]
		&\quad x_{\ell,k,j} \!\in\! \{0,\,1\} \quad \forall \ell, \, \forall k, \forall j= 1, \ldots, T_k \, ,  
	\end{align}
\end{subequations}
where $\dot{f}_{\ell}$ denotes the computational capacity of the MEC server $\ell$, and constraint~\eqref{prob:knap:C2}, originally with equality, has been relaxed.
In the above problem, the \textit{knapsacks} are the MEC servers at the APs and at the CPU, the knapsacks' capacity is the number of CPU cycles available at each MEC server, the objects to be put in the knapsacks are the computational loads to be offloaded, and, finally, the \textit{weight} (and \textit{profit}) of the generic computational task is given by $\xi_{k,j}(t,\bp)$.  
Problem \eqref{prob:knap} can be solved using some of the methods available in the literature, see for instance~\cite[Chapter 6]{martello1990knapsack}.
Once the problem has been solved, the values of the output variables in $\mathcal{X}$ provide the sought allocation of the CPU cycles to the tasks. For instance, if, for certain values $t\!=\!t^*$, $\ell\!=\!\ell^*$, $k\!=\!k^*$ and $j\!=\!j^*$ we have $x_{\ell^*,k^*,j^*}\!=\!1$, this means that subtask $j^*$ of user $k^*$ is to be executed at the MEC server $\ell^*$, using $\xi_{k^*,j^*}(t^*,\bp)$ cycles per second. Hence, the optimal $\mathcal{F}$ is obtained upon the optimal $\mathcal{X}$ and value of $t$ at the last iteration of the bisection method as $\mathcal{F} \!=\!\left\{ \xi_{k,j}(t,\bp)\,x_{\ell,k,j}: \, k\!=\!1, \ldots, K; \, \ell\in\mathcal{G}_k; \, j\!=\!1, \ldots, T_k \right\}.$

\subsubsection{Optimization with respect to $\bp$ and $\bnu$}
Let us assume now that the elements in $\mathcal{F}$ are fixed and let us solve problem \eqref{prob:P1} with respect to $\bp$ and $\bnu$. Basically, we have to minimize the objective function in  \eqref{prob:P1} with respect to $\bp$ and $\bnu$ and with constraints 
\eqref{prob:P1:C1}, \eqref{prob:P1:C2} and \eqref{prob:P1:C6}. The problem is not convex due to the spectral efficiency expression $\text{SE}_k(\bp)$ in constraints \eqref{prob:P1:C1} and  \eqref{prob:P1:C2}.
We thus resort to \textit{sequential convex programming}, that is an iterative optimization framework wherein in each iteration we optimize a related convex approximation of the original problem. 
Notice that the uplink instantaneous SE for user $k$ can be expressed as
\begin{align} \label{eq:SE:concavity}
&\text{SE}_k(\bp) \!=\! \frac{\tu}{\tc}\log_2 \left(1\!+\!\frac{\mathsf{num}_k(\bp)}{\mathsf{den}_k(\bp)}\right) \nonumber \\
&\quad\!=\! \frac{\tu}{\tc}[\log_2 (\mathsf{num}_k(\bp)\!+\!\mathsf{den}_k(\bp))\!-\!\log_2 (\mathsf{den}_k(\bp))],
\end{align}    
where $\mathsf{num}_k(\bp)$ and $\mathsf{den}_k(\bp)$ describe the numerator and the denominator of~\eqref{eq:SINR}, respectively, which are functions of the power coefficients.
As $\log_2(\cdot)$ is increasing and the summation preserves concavity, the r.h.s. of~\eqref{eq:SE:concavity} is the difference of two concave functions. Recalling that any concave function is upper-bounded by its Taylor expansion around any given point $\bp^{(0)}$, a concave lower-bound of $\text{SE}_k(\bp)$ is obtained as
\begin{align} \label{eq:SE-lower-bound}
\text{SE}_k(\bp) &\geq \frac{\tu}{\tc}\bigg[ \log_2 (\mathsf{num}_k(\bp)\!+\!\mathsf{den}_k(\bp))\! -\! \log_2 (\mathsf{den}_k(\bp^{(0)})) \nonumber \\
&\qquad\!-\! \nabla\trans_{\bp} \log_2(\mathsf{den}_k(\bp))\Big\rvert_{\bp = \bp^{(0)}} (\bp \!-\! \bp^{(0)})\bigg] \nonumber \\
&= \widetilde{\text{SE}}_k(\bp,\bp^{(0)})\,.
\end{align}
Hence, constraints~\eqref{prob:P1:C1} and~\eqref{prob:P1:C2} can be approximated and convexified by taking
\begin{align*}
\frac{b_k}{B~\widetilde{\text{SE}}_k(\bp,\bp^{(0)})} + \displaystyle \max_{j=1, \ldots, T_k}\frac{w_{k,j}}{f_k(j)} &\leq \mathcal{\widetilde{L}}_k \, ,~\forall k, \\
\widetilde{\text{SE}}_k(\bp,\bp^{(0)}) &\geq \nu_k\, ,~\forall k,
\end{align*}
for any feasible choice of $\bp^{(0)}$.
The arguments above hold if the receive combining vector is independent of the uplink powers. This is not true in general, as for the P-MMSE combining scheme in~\eqref{eq:PMMSE-combining-vector}. 
In this case, $\widetilde{\text{SE}}_k(\bp,\bp^{(0)})$ is still a non-linear function of the uplink powers.  
This issue can be tackled by treating the combining vectors at the $n$-th iteration of the SCA optimization framework as constant with respect to the current uplink transmit powers $\bp^{(n)}$, and being exclusively function of $\bp^{(n-1)}$.
Hence, the problem to be solved at the $n$-th iteration of the proposed SCA method can be formulated as
\begin{subequations} \label{prob:P2}
\begin{align}	
  \mathop {\text{minimize}}\limits_{\bp^{(n)},~\bnu^{(n)}} & \;\; \vpp \boldsymbol{1}\trans_K \bp^{(n)} - \vpse \boldsymbol{1}\trans_K \bnu^{(n)} \label{prob:P2:obj} \\[-1ex]
  \text{s.t.} &\;\; \frac{b_k}{B~\widetilde{\text{SE}}_k(\bp^{(n)},\bp^{(n-1)})\big\vert_{\bv_k^{(n)}(\bp^{(n-1)})}} \nonumber \\
  &\;\;\qquad+ \max_{j=1, \ldots, T_k} \frac{w_{k,j}}{f_k(j)} \leq \mathcal{\widetilde{L}}_k, \forall k, \label{prob:P2:C1} \\[0ex]
  			  &\;\; \widetilde{\text{SE}}_k(\bp^{(n)},\bp^{(n-1)})\big\vert_{\bv_k^{(n)}(\bp^{(n-1)})} \geq \nu^{(n)}_k,\,\forall k,\label{prob:P2:C2} \\[-.5ex]
  			  &\;\; \bzero_K \preceq \bp^{(n)} \preceq p_{\text{max}}\cdot \boldsymbol{1}_K, \label{prob:P2:C5} 
\end{align}
\end{subequations}
where $\vpp$ and $\vpse$ are set as in~\eqref{eq:weights} with $\text{SE}^{(0)}_k \!=\! \text{SE}_k(\bp^{(0)}), \, \forall k$.
Notice that the notation $\widetilde{\text{SE}}_k(\bp^{(n)},\bp^{(n-1)})\big\vert_{\bv_k^{(n)}(\bp^{(n-1)})}$ emphasizes that the receive combining vectors, involved in the expression of the SE, are constant with respect to the current transmit powers $\bp^{(n)}$ (i.e., the optimization variables), and are computed upon the optimal values of the transmit powers at the previous iteration of the SCA algorithm, namely $\bp^{(n-1)}$.
For any iteration $n$ of the SCA method, $\widetilde{\text{SE}}_k(\bp^{(n)}, \bp^{(n-1)})\big\vert_{\bv_k^{(n)}(\bp^{(n-1)})}$ is a suitable convex approximation of $\text{SE}_k(\bp^{(n)})$, as  
the following properties are fulfilled~\cite{Marks1978}:
\begin{subequations}
\begin{align}
&\text{SE}_k(\bp^{(n)}) \!\geq\! \widetilde{\text{SE}}_k(\bp^{(n)}, \bp^{(n-1)})\big\vert_{\bv_k^{(n)}(\bp^{(n-1)})}, \, \forall n,\,\forall k, \label{eq:sca:prop1} \\[-.5ex] 
&\text{SE}_k(\bp^{(n-1)}) \!=\! \widetilde{\text{SE}}_k(\bp^{(n-1)}, \bp^{(n-1)}), \, \forall n,\,\forall k, \label{eq:sca:prop2} \\[-.5ex]  
&\nabla_{\bp} \text{SE}_k(\bp^{(n-1)}) \!=\! \nabla_{\bp} \widetilde{\text{SE}}_k(\bp^{(n-1)}, \bp^{(n-1)}), \, \forall n,\,\forall k. \label{eq:sca:prop3} 
\end{align} 
\end{subequations}
According to the theory in~\cite{Marks1978}, by virtue of the properties~\eqref{eq:sca:prop1},~\eqref{eq:sca:prop2}, the sequence of the values attained by the objective function~\eqref{prob:P1:obj} at the optimal points of
each iteration of the SCA algorithm is monotonically decreasing with the increase of the iteration number and converges to a finite limit. Moreover, due to the property~\eqref{eq:sca:prop3}, the optimal solution of the SCA algorithm at convergence satisfies the Karush-Kuhn-Tucker (KKT) conditions of problem~\eqref{prob:P1}.

Algorithm~\ref{alg:SCA}, to be run at the CPU, summarizes the proposed alternate maximization strategy for  sub-optimally solving problem~\eqref{prob:P1}.
\setlength{\textfloatsep}{5mm}
\begin{algorithm}[!t]
	\small
	\setstretch{1}
	\caption{Alternating optimization for problem~\eqref{prob:P1}}
	\vspace{1mm}
	\textbf{Input:} Any choice of feasible transmit powers $\bp^{(0)}$, $M_{\max}$, $\epsilon$\,;
	\begin{algorithmic}[1]
		\State Compute the SE values:  $\bnu^{(0)}
		\gets \left[\text{SE}_1(\bp^{(0)}), \ldots, \text{SE}_K(\bp^{(0)})\right]\trans $;
		\State Evaluate the objective function in \eqref{prob:P1};
		\State Initialize $m \gets 1$;
		\Repeat
		\Statex $\hspace*{5mm}$\texttt{\%\%  Updating $\mathcal{F}$}
		\State Check feasibility of \eqref{prob:knap} with $t=+\infty$;
		\If {\eqref{prob:knap} is unfeasible} 
		\State {Exit procedure and declare the problem unfeasible;}
		\Else 
		\State Set $t_0$ and $t_1$ as in \eqref{eq:bisection_interval};
		\Repeat {\texttt{~\%\%  Bisection algorithm}}
		\State  $t \gets (t_0 + t_1)/2$;
		\State Solve problem \eqref{prob:knap} with current value of $t$;
		\IfThenElse{\eqref{prob:knap} is unfeasible}{$t_0 \gets t$;}{$t_1 \gets t$;}
		\Until $|(t_1 -t_0)/t_1| \leq \epsilon$
		\State Set $\mathcal{F}$ as resulting from the solution of  \eqref{prob:knap} with $t=t_1$;
		\Statex $\hspace*{10mm}$\texttt{\%\% Updating $\bp$ and $\bnu$}
		\State Initialize $n \!\gets\! 1$; $\widetilde{\bp}^{(0)}\!\gets\!\bp^{(n)}$;
		\State Initialize $\widetilde{\bnu}^{(0)} \!\gets\! [\text{SE}_1(\widetilde{\bp}^{(0)}), \ldots, \text{SE}_K(\widetilde{\bp}^{(0)})]\trans$;
		       \Repeat  {\texttt{~\%\%  SCA algorithm}}
		       		\State Let $\bp^{\star}, \bnu^{\star}$ be the optimal solutions of problem~\eqref{prob:P2};
					\State $\widetilde{\bp}^{(n)} \gets \bp^{\star}$; $\widetilde{\bnu}^{(n)} \gets \bnu^{\star}$; $n \gets n+1$;
				\Until convergence
		\EndIf
		\State $m \gets m + 1$;
		\State Evaluate the new value of the objective function in \eqref{prob:P1};
		\Until the objective function in \eqref{prob:P1} converges or $m == M_{\max}$
	\end{algorithmic}
	\textbf{Output:} $\bp, \mathcal{F}, \bnu$;
	\label{alg:SCA}
\end{algorithm}

\subsection{Algorithm initialization} \label{subsec:feasibility}
We now discuss on how to initialize with a suitable power vector $\bp^{(0)}$ the alternating optimization procedure in Algorithm~\ref{alg:SCA}. The procedure that we adopt is the following: 
first of all, we start by considering an allocation of the computational resources in $\mathcal{F}$ minimizing the maximum SE requirement, so that the load of the latency constraint on the air interface of the system is minimized; next, we compute the transmit powers needed to achieve the found minimum SE requirement for all the users. The obtained values of the transmit power will be used to initialize the proposed JCPA algorithm.
To begin with, we notice that constraint \eqref{prob:P1:C1} can be written as
\begin{equation}
	\text{SE}_k(\bp) \geq
	\frac{b_k/B}{\mathcal{\widetilde{L}}_k - \max\limits_{j=1, \ldots, T_k}\dfrac{w_{k,j}}{f_k(j)}}\, , \quad \forall k \; .
\end{equation}
We thus seek for a set of computational rates $\mathcal{F}$ minimizing the maximum SE requirement, i.e. we focus on the optimization problem  
\begin{align}	\label{prob:P4:feasibility_f}
\begin{split}
  \mathop {\text{minimize}}\limits_{\mathcal{F}} &\quad\max\limits_k \frac{b_k/B}{\mathcal{\widetilde{L}}_k - \max\limits_{j=1, \ldots, T_k} \dfrac{w_{k,j}}{f_k(j)}}\\ \text{s.t.} &\quad \eqref{prob:P1:C3}, \eqref{prob:P1:C4}, \eqref{prob:P1:C5}, \eqref{prob:P1:C7} \; , 			  
  \end{split}
\end{align} 
which can be written in epigraph form as
\begin{subequations} \label{prob:P5:feasibility_f}
\begin{align}	
 \mathop {\text{min}}\limits_{\mathcal{F}} &\;\; t \\
 \text{s.t.} &\;\; f^{\mathsf{CPU}}_k(j) \!+\! \sum_{l \in \mathcal{O}_k} f^{\mathsf{AP}}_{l,k}(j) \!\geq\! \displaystyle \frac{w_{k,j}}{\mathcal{\widetilde{L}}_k-\dfrac{b_k}{t\, B}}, \forall k, \label{prob:P5:C1} \\[-.5ex]
 &\; \eqref{prob:P1:C3}, \eqref{prob:P1:C4}, \eqref{prob:P1:C5}, \eqref{prob:P1:C7}\, . \label{prob:P5:C2}
\end{align}
\end{subequations}  
The above problem has the same structure as \eqref{prob:P1_F2} and can be solved following the same procedure outlined in the previous subsection.
Let us denote by $\mathcal{F}^{\star}$ the solution to problem \eqref{prob:P5:feasibility_f}. We are now ready to compute the transmit vectors that fulfill the minimum SE requirement. Assuming that the computational rates in \eqref{eq:latency-constraint} are those corresponding to $\mathcal{F}^{\star}$, the latency constraint requires that 
\begin{equation} \label{eq:latency-constraint-2}
 R_k(\bp) \geq \frac{b_k}{\mathcal{\widetilde{L}}_k - \max\limits_{j=1, \ldots, T_k} \dfrac{w_{k,j}}{f^{\star}_k(j)}     }, \ \forall k,
\end{equation}
which represents a QoS requirement. The above  inequality translates to the following instantaneous SINR requirement
\begin{equation} \label{eq:SINR-requirement}
\frac{p_k g_{kk}}{\sum\limits_{i \neq k}^K p_i \left( g_{ki} + c_{ki} \right) + p_k c_{kk} + \sigma^2 \norm{\bv_{k}\herm \bD_k}^2} \geq \underbracket[.5pt]{2^{z_k} - 1}_{\widetilde{\gamma}_k}, \ \forall k,
\end{equation}
where $g_{ki} = |\bv_{k}\herm \bD_{k} \hat{\bh}_i|^2$, $c_{ki} = \bv_{k}\herm \bD_k  \bC_i \bD_k \bv_{k}$, and $$z_k = \dfrac{b_k \tc }{B\tu}{\left( \mathcal{\widetilde{L}}_k - \max\limits_{j=1, \ldots, T_k}\dfrac{w_{k,j}}{f^{\star}_k(j)} \right)}^{\!-1}.$$
Hence, the instantaneous SINR requirement in~\eqref{eq:SINR-requirement} can be rewritten as the vector inequality
\begin{equation} \label{eq:powers-requirement}
\bp \succeq \boldsymbol{\Upsilon}\bG^{-1}(\bZ \bp + \sigma^2 \bu),
\end{equation}
where $\boldsymbol{\Upsilon} \!\!=\! \text{diag}(\widetilde{\gamma}_1, \ldots, \widetilde{\gamma}_K)$, $\bu \!=\! \left[\norm{\bv_{1}\herm \bD_1}^2 \!\ldots \norm{\bv_{K}\herm \bD_K}^2 \right]\trans$, and
\begin{align}
\begin{split}
[\bG]_{ki} &= \begin{cases}
g_{kk} - c_{kk} \widetilde{\gamma}_{k}, & \text{if } k = i, \\[-.5ex]
0, & \text{otherwise,}
\end{cases} \\
[\bZ]_{ki} &= \begin{cases}
0, & \text{if } k = i, \\[-.5ex]
g_{ki} + c_{ki}, & \text{otherwise.}
\end{cases} \label{eq:standard-algorithm}
\end{split}
\end{align}%
\vspace*{-3mm}
\begin{algorithm}[!t]
\small
\setstretch{1}
\caption{\textit{Standard} power control assuming P-MMSE}
\vspace{1mm}
\textbf{Input:} $\{ \boldsymbol{\Upsilon} \}$, $\{\bC_k \}$, $\{\bD_k \}$, $\{\hat{\bh}_k \}$;
\begin{algorithmic}[1]
\State Initialize $n = 0$, $\chi = 1$; $\bp^{(0)}$; $\bv(\bp^{(0)})$;
\While {$\chi > 0.005$}	
	\State Compute $\bG(\bp^{(n)})$, $\bZ(\bp^{(n)})$ according to~\eqref{eq:standard-algorithm};
	\State Compute $\rho(\boldsymbol{\Upsilon}\bG^{-1}\bZ)$;
	\If {$\bG$ is non-negative \textbf{and} $\rho<1$ }
		\State $n \leftarrow n+1$;
		\State $\bp^{(n)} \leftarrow \bI(\bp^{(n-1)})$;
		\State Compute $\bv_k(\bp^{(n)})$ according to~\eqref{eq:PMMSE-combining-vector}, $\forall k$;
		\State $\chi \leftarrow \max\limits_k \left| \dfrac{p^{(n)}_k - p^{(n-1)}_k}{p^{(n-1)}_k} \right|$;
	\Else { {Exit procedure and declare the problem unfeasible;}}  
	\EndIf	
\EndWhile
\end{algorithmic}
\textbf{Output:} $\bp^{\star} \leftarrow \bp^{(n)}$;
\label{alg:power-control}
\end{algorithm}
  
Hence, a set of nonnegative uplink powers can be determined capitalizing on the requirement in~\eqref{eq:powers-requirement}. The set of SINR targets $\{\widetilde{\gamma}_{k}\}$ is feasible if and only if all the diagonal elements of $\bG$ are nonnegative\footnote{Notice that it is not guaranteed that $g_{kk} - c_{kk} \widetilde{\gamma}_{k} \geq 0, \forall k.$}, and the Perron-Frobenius eigenvalue of the matrix $\boldsymbol{\Upsilon}\bG^{-1} \bZ$, denoted by $\rho$, is real and nonnegative, and $\rho < 1$~\cite{Ulukus1998}.  
If these conditions are satisfied, then $\bI(\bp) = \boldsymbol{\Upsilon}\bG^{-1}(\bZ \bp + \sigma^2 \bu)$ is a \textit{standard interference function}~\cite{Yates1995}, and an optimal solution for the uplink transmit powers, is obtained iteratively through the \textit{standard power control algorithm}~\cite{Yates1995} as $\bp^{(n)} = \bI(\bp^{(n-1)})$, for any given initial choice $\bp^{(0)}$.\footnote{This optimal solution satisfies all the inequalities in~\eqref{eq:powers-requirement} with equality, and minimizes the sum of the transmitted powers~\cite{Yates1995}.} Algorithm~\ref{alg:power-control} specifies the steps of the \textit{standard power control algorithm} based on sequential optimization, and assuming P-MMSE. If Algorithm~\ref{alg:power-control} converges to an optimal solution, say $\bp^{\star}$, then this solution can be used as initial feasible choice for Algorithm~\ref{alg:SCA}, that is $\bp^{(0)} = \bp^{\star}$. Whenever a set of feasible uplink transmit powers is found (i.e., line 7 of Algorithm~\ref{alg:power-control}), the receive combining vectors $\{\bv_k\}$ must be updated accordingly, such that the matrix $\boldsymbol{\Upsilon}\bG^{-1} \bZ$ at the next iteration is properly computed.
If this two-stage procedure fails to find a feasible set of uplink powers and computational rates, then a non-empty feasible set for problem~\eqref{prob:P2} can anyhow be enforced by a proper admission control~\cite{WChen2019,JZheng2019}.

\section{Benchmarks}\label{sec:benchmarks}  
\subsection{Co-located Massive MIMO System Model and Resource Allocation}
\label{subsec:cellular}
A co-located massive MIMO system can be seen as a special case of a CF-mMIMO wherein each user is served by only one of the few deployed base stations (BSs), each of which is equipped with many antennas. An achievable uplink SE for user $k$ served by BS $l$, is given by
\begin{align} 
\overline{\text{SE}}^{(\text{cell})}_{lk} &\!=\! \frac{\tu}{\tc} \EX{\log_2 (1 + \text{SINR}^{(\text{cell})}_{lk})}, \quad \text{where} \label{eq:SE-cellular}\\
\text{SINR}^{(\text{cell})}_{lk} &\!=\! \frac{p_k |\bv_{lk}\herm \hat{\bh}_{lk}|^2}{\sum\limits_{i \neq k}^K p_i |\bv_{lk}\herm \hat{\bh}_{li}|^2 \!+\! \bv_{lk}\herm\! \left( \sum\limits_{i=1}^K p_i \bC_{li} \right)\! \bv_{lk} \!+\! \sigma^2 \norm{\bv_{lk}}^2},  \label{eq:SINR-cellular}
\end{align}
for an arbitrary receive combining vector $\bv_{lk} \!\in\! \C^M$. The effective uplink SINR, $\text{SINR}^{(\text{cell})}_{lk}$, is maximized by using the \textit{Local}-MMSE\footnote{The AP in each cell performs MMSE combining on its own, only relying upon the local channel estimates.} (L-MMSE) receive combining, which is
\begin{equation} \label{eq:LMMSE-combining-vector}
\bv_{lk}^{\mathsf{L-MMSE}} = p_k \left( \sum\limits_{i=1}^K p_i \big( \hat{\bh}_{li} \hat{\bh}_{li}\herm \!+\! \bC_{li} \big) \!+\! \sigma^2 \bI_{M} \right)^{\!\!\!-1} \hat{\bh}_{lk}.
\end{equation}
L-MMSE is not scalable but can be used as a benchmark, since it constitutes the optimal combining scheme for cellular networks~\cite{cellfreebook}.
For co-located massive MIMO, we can reasonably assume that only the serving BS offers computational facility to its user terminals.
Hence, using the same notation as in~\eqref{prob:P1}, we can formulate the JPCA problem for cellular networks as
\begin{subequations} \label{prob:P1:cellular}
\begin{align}	
  \mathop {\text{minimize}}\limits_{\bp,~\boldsymbol{\zeta}\in \mathbb{R}_{>0}^{K},~\bnu} & \quad \vpp \boldsymbol{1}\trans_K \bp - \vpse \boldsymbol{1}\trans_K \bnu  \label{prob:P1:cellular:obj} \\[-1ex]
  \text{s.t.} &\quad \frac{b_k}{B\,\text{SE}_{lk}(\bp)} \!+\! \frac{w_k}{\boldsymbol{\zeta}(k)} \!\leq\! \mathcal{L}^{\text{cell}}_k,~\forall k,\, l \in \mathcal{M}_k, \label{prob:P1:cellular:C1} \\[-0ex]
  			  &\quad \text{SE}_{lk}(\bp) \geq \nu_{k},~\forall k,\, l \in \mathcal{M}_k,\label{prob:P1:cellular:C2} \\[-0ex]
  			  &\quad \tilde{\bc}_l\trans \boldsymbol{\zeta} \leq f^{\mathsf{BS}}_l, \forall l, \label{prob:P1:cellular:C3} \\[-0ex]
  			  &\quad \bzero_K \preceq \bp \preceq p_{\text{max}}\cdot \boldsymbol{1}_K, \label{prob:P1:cellular:C4}
\end{align}
\end{subequations}
where $\text{SE}_{lk}$ represents the instantaneous SE, that is the value attained by~\eqref{eq:SE-cellular} without expectation and $\mathcal{L}^{\text{cell}}_k$ denotes the maximum tolerable latency for user $k$ in the cellular setup. This latency value is presumably larger than its cell-free counterpart as it does not include the delay produced by the fronthaul signalling (i.e., steps 2--5 in~\Figref{fig:signalling-diagram}).
Moreover, in~\eqref{prob:P1:cellular}, $f^{\mathsf{BS}}_l$ indicates the computational capability of BS $l$, $\boldsymbol{\zeta}$ denotes a $K\!\times\!1$ vector of optimization variables, where its $k$-th element, $\boldsymbol{\zeta}(k)$, represents the amount of computational resources allocated to user $k$ by its serving BS $l$, that is $\boldsymbol{\zeta}(k) \!=\! f^{\mathsf{BS}}_{l,k},~l\!\in\! \mathcal{M}_k$, with $|\mathcal{M}_k| \!=\! 1$. Moreover, $\tilde{\bc}_l\! \in\! \{0,1\}^K$ is an auxiliary binary vector, where the $k$-th element is 1 if BS $l$ serves user $k$, and 0 otherwise. 
Lastly, $\nu_{k}$ represents the minimum instantaneous SE for user $k$.
Problem~\eqref{prob:P1:cellular} can be convexified via sequential optimization, using a similar methodology as in~\eqref{eq:SE-lower-bound}.
Hence, the optimization problem at the $n$-th iteration of the SCA method can be formulated as
\begin{subequations} \label{prob:P2:cellular}
\begin{align}	
  \mathop {\text{minimize}}\limits_{\substack{\bp^{(n)},\,\bnu^{(n)} \\\boldsymbol{\zeta}^{(n)}\in \mathbb{R}_{>0}^{K}}} & \quad \vpp \boldsymbol{1}\trans_K \bp^{(n)} - \vpse \boldsymbol{1}\trans_K \bnu^{(n)}  \label{prob:P2:cellular:obj} \\[-.5ex] 	
  \text{s.t.} &\quad \frac{b_k}{B~\widetilde{\text{SE}}_{lk}(\bp^{(n)},\bp^{(n-1)})\big\vert_{\bv_{lk}^{(n)}(\bp^{(n-1)})}} + \frac{w_k}{\boldsymbol{\zeta}^{(n)}(k)} \nonumber \\[-0.5ex]
  &\quad\qquad \leq \mathcal{L}^{\text{cell}}_k,~\forall k, \forall l \in \mathcal{M}_k, \label{prob:P2:cellular:C1} \\[-0ex]
  			  &\quad \widetilde{\text{SE}}_{lk}(\bp^{(n)},\bp^{(n-1)})\big\vert_{\bv_{lk}^{(n)}(\bp^{(n-1)})} \geq \nu^{(n)}_k, \nonumber \\[-0.5ex]
  			  &\quad\qquad \forall k, \forall l \in \mathcal{M}_k, \label{prob:P2:cellular:C2} \\[-0ex]
  			  &\quad \tilde{\bc}_l\trans \boldsymbol{\zeta}^{(n)} \leq f^{\mathsf{BS}}_l, \forall l, \label{prob:P2:cellular:C3} \\[-0ex]
  			  &\quad \bzero_K \preceq \bp^{(n)} \preceq p_{\text{max}}\cdot \boldsymbol{1}_K, \label{prob:P2:cellular:C4}
\end{align}
\end{subequations}
where $\vpp$ is set as in~\eqref{eq:weights}, $\vpse = \ose/[K \max\limits_{l,k} \text{SE}_{lk}(\bp^{(0)})]$, and $\widetilde{\text{SE}}_{lk}(\bp^{(n)},\bp^{(n-1)})$ is a concave lower-bound of $\text{SE}_{lk}(\bp^{(n)})$ around the point $\bp^{(n-1)}$, obtained by using the same methodology as in~\eqref{eq:SE-lower-bound}.
The SCA algorithm is run in a centralized fashion by a network entity, e.g., one of the~BSs, an its
convergence is guaranteed, as $\widetilde{\text{SE}}_{lk}(\bp^{(n)},\bp^{(n-1)})$ is a suitable convex approximation of $\text{SE}_{lk}(\bp^{(n)})$.
As per the feasibility, problem~\eqref{prob:P2:cellular} admits a non-empty feasible set if
\begin{align}
\label{eq:nonempty-set:cellular}
R_k >\frac{b_k}{\mathcal{L}^{\text{cell}}_k},\, \forall k, \;\text{and}\;
\sum\limits_{k \in \mathcal{K}_l} \dfrac{w_k}{\mathcal{L}^{\text{cell}}_k - b_k/R_{lk}} < f^{\mathsf{BS}}_l,~\forall l,
\end{align}
where $R_{lk} \!=\! B\!\times\! \text{SE}_{lk}$, is the uplink instantaneous rate of user $k$ served by BS $l$, and $\mathcal{K}_l$ is the set of the users served by BS $l$.
The conditions in~\eqref{eq:nonempty-set:cellular} are necessary but not sufficient due to the interference-limited scenario that makes simultaneously maximizing the per-user SEs intractable.

\subsection{Heuristic Resource Allocation for Distributed Network Topology}
\label{subsec:C-RAN}
In this section, we propose an alternative approach to the JPCA for cell-free massive MIMO which consists in heuristically allocating the MEC server computational resources to the users according to a pre-determined metric, and then optimizing with respect to $\bp$ and $\bnu$ as described in~\Secref{subsec:JPCA}. Hence, unlike the JPCA, such a heuristic resource allocation does not jointly optimize the uplink power consumption and the allocated computational resources. However, it represents a low-complexity solution with respect to the JPCA which requires solving a multiple knapsack problem---an \textit{NP-hard} problem in strong sense~\cite{martello1990knapsack}. The psuedo-code of the proposed heuristic allocation of uplink powers and computational resources is reported in Algorithm~\ref{alg:cran-scheme}.
As for this heuristic scheme, we assume that a subtask offloaded by any user can either be processed at the CPU MEC server or at one of the AP MEC servers. 
Firstly, the subtasks are sorted in descending order by the metric $\mu_{k,j} = w_{k,j} {\left(\widetilde{\mathcal{L}}_k-{b_k}/{R_k} \right)}^{-1} \, [\text{cycles/s}],~k=1, \ldots, K; i=1, \ldots, T_k,$ which represents the computational demand of user $k$ for subtask $j$ related to its latency requirements and uplink rate, conditioned to a pre-determined power allocation. Then, each task is offloaded at the MEC server with more available computing resources, either at one of the APs or at the CPU. The fractions of computational resources assigned to each user's subtask are further scaled so as to saturate the computational capabilities of those APs and (possibly) the CPU involved in the offloading process. Once the set $\mathcal{F}$ is determined, Algorithm~\ref{alg:cran-scheme} concludes by solving problem~\eqref{prob:P2} as described in~\Secref{subsec:JPCA}. Hence, this scheme heuristically allocates the remote computational resources and only optimizes with respect to $\bp$ and $\boldsymbol{\nu}$, unless the computational capabilities at the MEC servers are insufficient, i.e., there exists at least a subtask $j$ of user $k$ such that $\mu_{k,j} \!>\! \max\limits_{\ell \in \mathcal{G}_k} \bar{f}_{\ell}$, with $\bar{f}_{\ell}$ being the ``online'' available computing resources at the CPU and at each of the AP MEC server.       
Notice that if the computational resource allocation problem in Algorithm~\ref{alg:cran-scheme} is feasible, then the necessary but not sufficient condition for the feasibility of problem~\eqref{prob:P2} at the first iteration is $R_k \!>\! b_k / \mathcal{\widetilde{L}}_k, \forall k$.
\begin{algorithm}[!t] 
\small
\setstretch{1.1}
\caption{Heuristic allocation of uplink powers and computational resources}
\vspace{1mm}
\textbf{Input:} $\{\mu_{k,j}\}$, $\{\dot{f}_{\ell}\}$;
\begin{algorithmic}[1]
\State Initialize $\dot{f}_{\ell,k}(j)=0,~\forall \ell,~\forall k,~\forall j$; 
\For{each task in descending order by $\mu_{k,j}$}
	\State $\bar{f}_{\ell} = \dot{f}_{\ell}-\sum\nolimits_{k=1}^K \sum\nolimits_{j=1}^{T_k} \dot{f}_{\ell,k}(j),~\ell \in \mathcal{G}_k$;
	\State $\ell^{\ast} =  \argmax\nolimits_{\ell} \bar{f}_{\ell}$;
	\If{$\mu_{k,j} \leq \bar{f}_{{\ell}^{\ast}}$} {$\dot{f}_{{\ell}^{\ast},k}(j) = \mu_{k,j}$;}
	\Else { Exit procedure and declare the problem unfeasible;}
	\EndIf
\EndFor
\State $\vartheta_{\ell} = \dot{f}_{\ell}/\sum\nolimits_{k=1}^K \sum\nolimits_{j=1}^{T_k} \dot{f}_{\ell,k}(j),~\forall \ell:~\exists~\ell,k,j.~\dot{f}_{\ell,k}(j) \neq 0$;
\State $\dot{f}_{\ell,k}(j) = \vartheta_{\ell} \dot{f}_{\ell,k}(j),~\forall \ell\!\in\!\mathcal{G}_k$;
		\State Initialize $n \gets 1$; $\widetilde{\bp}^{(0)}\gets\bp^{(n)}$;
		\State Initialize $\widetilde{\bnu}^{(0)} \gets [\text{SE}_1(\widetilde{\bp}^{(0)}), \ldots, \text{SE}_K(\widetilde{\bp}^{(0)})]\trans$;
		       \Repeat  {\texttt{~\%\%  SCA algorithm}}
		       		\State Let $\bp^{\star}, \bnu^{\star}$ be the optimal solutions of problem~\eqref{prob:P2};
					\State $\widetilde{\bp}^{(n)} \gets \bp^{\star}$; $\widetilde{\bnu}^{(n)} \gets \bnu^{\star}$; $n \gets n+1$;
				\Until convergence
	\end{algorithmic}
	\textbf{Output:} $\bp, \mathcal{F}, \bnu$;
	\label{alg:cran-scheme}	
\end{algorithm}

\subsection{Small-Cell Implementation and Resource Allocation}
\label{subsec:smallcell}  
With the terminology \textit{small-cell} we indicate an instance of cell-free network where each user receives communication service from only one of the APs and computational offloading service from either one AP or the CPU. With respect to an arbitrary user $k$, it holds $|\mathcal{M}_k| \!=\! 1$ and $|\mathcal{G}_k| \!=\! 1$, and not necessarily $\mathcal{M}_k$ coincides with $\mathcal{G}_k$. A similar MEC-enabled architecture was advocated in~\cite{Mukherjee2020} whose task offloading model is random and arbitrarily hinges on an offloading probability. Moreover, in~\cite{Mukherjee2020} both uplink transmit powers and allocated computational resources are fixed rather than optimized.    
Conversely, we herein consider a deterministic, heuristic task offloading model which accounts for the user computational demands and available remote computing resources.
Specifically, the set $\mathcal{M}_k$ comprises the AP with the best average channel gain towards user $k$. While, the set $\mathcal{G}_k$ comprises the MEC server with more available computing resources according to Algorithm~\ref{alg:cran-scheme} (up to line 15). Importantly, since each user receives computational offloading service from only one MEC server, the offloading process is carried out on a task basis rather than on a subtask basis, namely for an arbitrary user $k$ it holds $T_k \!=\! 1$.  
As for the small-cell implementation, an achievable uplink SE for user $k$ served by AP $l$ is given by~\eqrefs{eq:SE-cellular}{eq:SINR-cellular}, which is maximized by the LMMSE combining scheme in~\eqref{eq:LMMSE-combining-vector}. Since data decoding is performed locally at the AP, there is no need for the AP to forward the uplink data signal $\by$ to the CPU through the fronthaul network, but it can transmit the $b_k$ bits directly to the MEC server in charge of the offloading process. As $b_k/C_{\mathsf{FH}} \ll 2 b_k M \xi /C_{\mathsf{FH}}$, and the former is very small, we assume that $\mathcal{L}_k \!=\! \mathcal{\widetilde{L}}_k$ for the small-cell implementation.

\section{Simulation Results}\label{sec:simulation-results}
We consider a coverage area of 1 km$^2$ served by a total number of antennas $N \!=\! LM \!=\! 400$. For the co-located massive MIMO setup, we choose $L \!=\! 4$ BSs, equipped with $M \!=\! 100$ antennas each, and deployed as a regular grid with intersite distance equal to 500 m. For the CF-mMIMO setup, we select $L \!=\! 100$ APs,  equipped with $M \!=\! 4$ antennas each, and deployed as a regular grid with intersite distance equal to 100 m. For all the setups a wrap-around simulation technique is used to remove the edge effects of the (nominal) coverage~area.
All the systems operate at 2~GHz carrier frequency, over a communication bandwidth $B\!=\!20$ MHz. The receiver noise power is conventionally set to -94 dBm, while the maximum transmit power per user is $p_{\text{max}} \!=\! 100$~mW.
The TDD coherence block is $\tc \!=\! 200$ samples long, $\td=0$, and $\tp = 5$ samples is the uplink training duration.
All the setups serve the same set of $K = 20$ users, that are uniformly distributed at random over the coverage area.
A random realization of users' locations defines a network snapshot, and determines a set of large-scale fading coefficients. These are computed according to the 3GPP Urban Microcell model defined in \cite[Table B.1.2.1-1]{LTE2017}.
The channel correlation matrices $\{ \bR_{lk} \}$ are generated by using the popular \textit{local scattering}~\cite[Sec. 2.5.3]{cellfreebook} model assuming half-wavelength spaced ULAs, and jointly Gaussian angular distributions of the multipath components around the nominal azimuth and elevation angles. The random variations in the azimuth and elevation angles are assumed to be independent, and the corresponding angular standard deviations (ASDs) are equal to $15^{\circ}$, which represents strong spatial channel correlation.

As $\tp \!<\! K$, pilots are to be re-assigned across users. To this end, we resort to the joint pilot assignment and AP (BS)-user association described in~\cite[Sec. 5.4]{cellfreebook}, so as to ensure that users served by the same set of APs (same BS, for the co-located setup) are given orthogonal pilots.   Concerning the power control,  we assume that the initial choice for the feasible transmit powers of the SCA algorithm follows, for all the setups, a fractional power control strategy given by
\begin{equation}
\label{eq:fractional_power_control}
[\bp^{(0)}]_k = p_{\text{max}} \frac{\big(\sum\nolimits_{l \in \mathcal{M}_k} \beta_{kl} \big)^{-0.5}}{\max\nolimits_{i \in \mathcal{S}_k}\left(\sum\nolimits_{l \in \mathcal{M}_i} \beta_{il} \right)^{-0.5}}, \forall k.
\end{equation}

Concerning the computation-offloading and latency model, for the cell-free setup we assume $f^{\mathsf{CPU}} \!=\! 10^{10}$ cycles/s, while $f_l^{\mathsf{AP}}$ are uniformly distributed random integers from the interval $[2,~4]\times10^9$ cycles/s. The latency requirements are $\mathcal{L}_k \!=\! 0.2$ s $\forall k$. The fronthaul capacity is $C_{\mathsf{FH}} \!=\! 10$ Gbps, and the number of bits for quantization is set as $\xi \!=\! 16$. As per the co-located setup, we select $f_l^{\mathsf{BS}} \!=\! \left\lceil \big({\sum\nolimits_{1=l}^{L^{\mathsf{AP}}} f_l^{\mathsf{AP}} \!+\! f^{\mathsf{CPU}}}\big)/{L^{\mathsf{BS}}}  \right\rceil,$ where $L^{\mathsf{AP}}$ is the number of APs in the cell-free setup, while $L^{\mathsf{BS}}$ is the number of BSs in the co-located setup, that is 100 and 4, respectively.  This choice ensures the same amount of available computational resources over the simulation area for both the setups. Lastly, the latency requirements for the users in the co-located setup is $\mathcal{L}_k^{\text{cell}} \!=\! 0.3$ s $\forall k$. Common to all the setups, the computational bits, $\{b_k\}$, are uniformly distributed random integers from the interval $[1, 4]$ Mbits, and the number of computation cycles needed to run the task itself is set as a linear function of $b_k$, that is $w_k \!=\! \alpha \, b_k$, with $\alpha \!=\! 50$ cycles/bit~\cite{Pradhan2020}. Finally, the number of subtasks any user's task is divided into is an integer drawn uniformly at random from the interval $[1, 4]$. An instance of the multiple knapsack problem, involved in Algorithm~\ref{alg:SCA}, is sub-optimally solved in polynomial time (with respect to the total number of subtasks) by using the \textit{Lagrangian Relaxation} technique~\cite[Section 6.2.2]{martello1990knapsack} combined with the \textit{cross-entropy} optimization method~\cite{rubinstein1999cross}.

\subsection{Performance Comparison between Network Architectures}
\label{subsec:performance-comparison}
\begin{figure}[!t]
	\centering
	\subfloat{\includegraphics[width=.8\columnwidth]{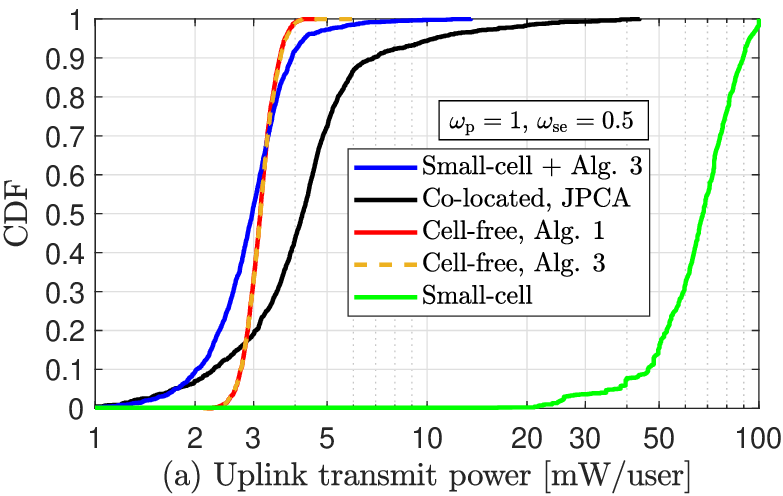}}\\[-0ex]
    \subfloat{\includegraphics[width=.8\columnwidth]{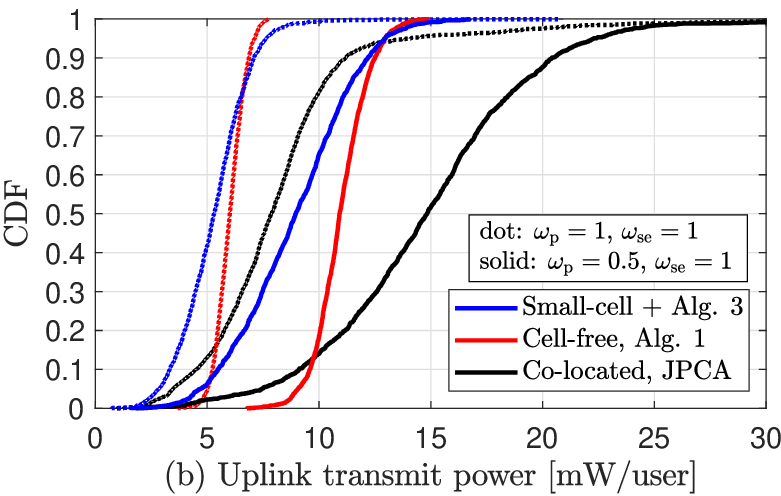}}\vspace*{-1.5ex}
	\caption{CDF of the per-user uplink transmit power for cell-free and co-located massive MIMO, assuming three configurations of $\{\op, \ose\}$. $K\!=\!20$, and $p_{\text{max}}\! =\! 100$ mW for all the users. The $x$-axis in~\Figref{fig:fig1}(a) is in logarithmic scale.}
	\label{fig:fig1}
\end{figure}

Firstly, we focus on the radiated power consumption. In~\Figref{fig:fig1}(a), we show the cumulative distribution function (CDF), obtained over 200 network snapshots, of the uplink transmit power per user, expressed in mWatt, being the solution of the Algorithm~\ref{alg:SCA} and the SCA problem~\eqref{prob:P2:cellular} for cell-free and co-located massive MIMO, respectively.
With the label ``Cell-free, Alg. 3'' we refer to the framework wherein uplink powers and computational rates are heuristically allocated according to Algorithm~\ref{alg:cran-scheme}.  
As for the small-cell implementation, we consider two cases: $(i)$ the label ``Small-cell'' indicates the framework wherein uplink powers and computational rates are not optimized, as in~\cite{Mukherjee2020}. The uplink powers result from~\eqref{eq:fractional_power_control}, while the computational rates are assigned according to the approach described in lines 1--15 of Algorithm~\ref{alg:cran-scheme}, with $T_k \!=\! 1$; $(ii)$ the label ``Small-cell + Alg.~\ref{alg:cran-scheme}'' indicates the framework wherein uplink powers and computational rates are heuristically allocated according to Algorithm~\ref{alg:cran-scheme}, with $T_k \!=\! 1$.In~\Figref{fig:fig1}(a), we consider the configuration:  $\op\!=\!1$, $\ose\!=\!0.5$, which applies to all the scheme but ``Small-cell''.
\\
\indent Numerical results reveal a dramatic transmit power saving for the CF-mMIMO users as compared to the co-located massive MIMO and the small-cell users.
Assuming $\op\!=\!1, \ose\!=\!0.5$,---which is the configuration that prioritizes the power saving over the transmission latency---at high percentiles, where the transmit power consumption is more significant, we indeed observe that the CF-mMIMO users can considerably reduce their transmit power as compared to the co-located massive MIMO and small-cell users. In co-located massive MIMO, those users with worse channel conditions, presumably at the cell-edge, need to employ more power to receive the required computational offloading service. In small-cell implementations instead, the users would benefit from a power optimization rather than employing a fixed power control strategy, such as fractional power control. To achieve excellent performance in small-cell implementations, the proposed Algorithm~\ref{alg:cran-scheme} should be employed for a proper resource allocation. \Figref{fig:fig1}(a) also highlights that the proposed heuristic allocation in Algorithm~\ref{alg:cran-scheme} performs as well as the proposed JPCA in Algorithm~\ref{alg:SCA}, in terms of power consumption.
At low percentiles, presumably corresponding to the users with better channel conditions and exiguous computational demands, the performance gap between co-located massive MIMO and CF-mMIMO (including its instance ``Small-cell + Alg.~\ref{alg:cran-scheme}'') reduces. Importantly, our JPCA scheme in CF-mMIMO is able to guarantee fairness among the users in terms of transmit power consumption.
In \Figref{fig:fig1}(b) we consider configurations giving equal and more weight to the SE with respect to the power consumption, with $\op\!=\!\ose\!=\!1$, and $(iii)$ $\op\!=\!0.5$, $\ose\!=\!1$, respectively. The levels of transmit power are larger than those attained by the previous configuration to guarantee larger SEs and thereby reducing the latency of the offloading process. Interestingly, the performance gap between CF-mMIMO and co-located massive MIMO increases as a higher SE is required. The macro-diversity gain enables CF-mMIMO to provide comparable SE levels to those of co-located massive MIMO, yet with lower uplink power consumption. 
\begin{figure}[!t]
	\centering
	\subfloat{\includegraphics[width=.85\columnwidth]{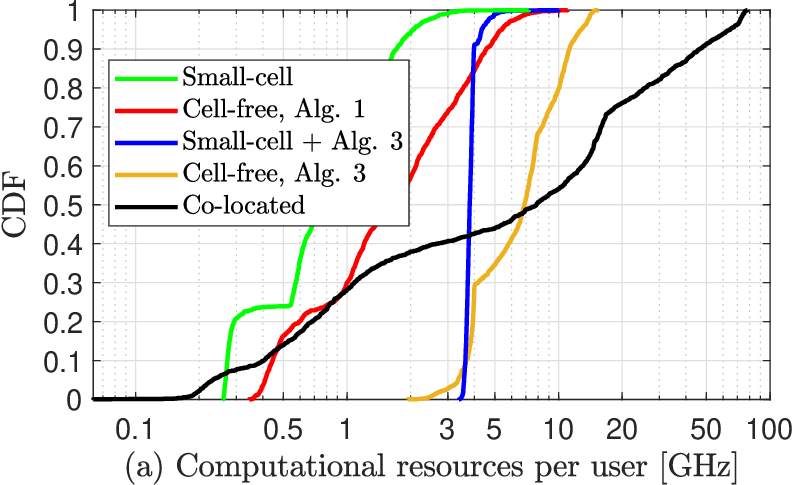}}\\[-0ex]%
	\subfloat{\includegraphics[width=.8\columnwidth]{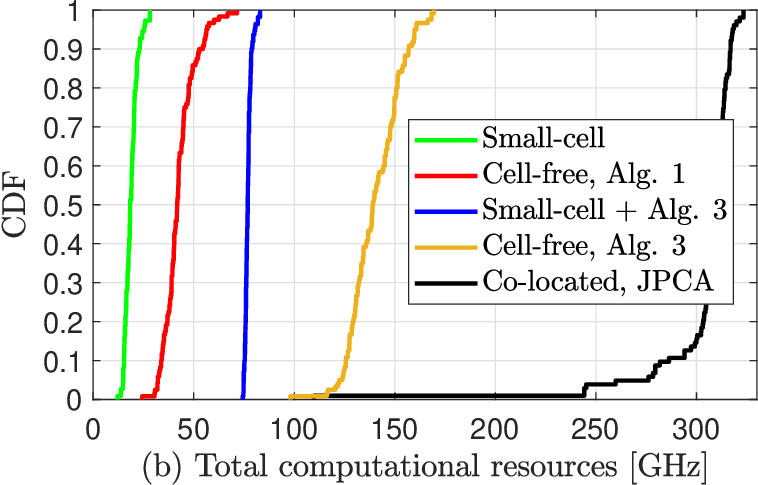}}\vspace*{-1.5ex} 
	\caption{CDF of the per-user (a) and total (b) computational resources allocated remotely, for cell-free, co-located massive MIMO and small-cell implementations. $f_l^{\mathsf{BS}}\!=\!\left\lceil \big({\sum\nolimits_{1=l}^{L^{\mathsf{AP}}} f_l^{\mathsf{AP}}\!+\!f^{\mathsf{CPU}}}\big)/{L^{\mathsf{BS}}}  \right\rceil$, $f^{\mathsf{CPU}}\!=\!10$ GHz, $f_l^{\mathsf{AP}}\!\!\sim\mathcal{U}(2,\,4)$ GHz, $\mathcal{L}_k = 200$ ms and $\mathcal{L}_k^{\text{cell}} = 300$ ms, $\forall k.$ The $x$-axis in~\Figref{fig:fig2}(a) is in logarithmic scale.}
	\label{fig:fig2}
\end{figure}\\
\indent To better motivate the previous performance, we now focus on the amount of computational resources allocated to the users, that is $\{f_k(i)\}$.
In \Figref{fig:fig2}(a), we show the CDF of the computational resources allocated to the single user, expressed in GHz ($10^9\times$cycles/s). While,~\Figref{fig:fig2}(b) shows the CDF of the total allocated computational resources, computed as $\sum_{k=1}^K\sum^{T_k}_{i=1} f_k(i)$. As we experienced negligible performance differences between the three considered weight configurations, we only report the results achieved with $\op\!=\!\ose\!=\!1$.~The amount of computing resources allocated by Algorithm~\ref{alg:cran-scheme} is significantly larger than that allocated by Algorithm~\ref{alg:SCA} for CF-mMIMO and small-cell implementations. Indeed, the fine-tuning of the allocated computational rates described by lines 16--19 of Algorithm~\ref{alg:cran-scheme} is carried out to utilize all the residual available resources after a first, conservative, feasible allocation based on the metric $\mu_{k,j}$ (the latter constitutes the ``Small-cell'' resource allocation approach). Allocating more computing resources entails reducing the computational latency and enables to increase the transmission latency as a result of lowering the uplink powers, while meeting the latency constraint. This motivates why ``Small-cell + Alg. 3'' allows higher levels of power saving than ``Cell-free, Alg. 1'', and how the performance gap, in terms of power consumption, between the nearly-optimal Algorithm~\ref{alg:SCA} and the heuristic Algorithm~\ref{alg:cran-scheme} is filled. Moreover, we recall that the effective latency constraint for the CF-mMIMO setup is stricter than that of its small-cell counterpart due to the fronthaul latency contribution, and this entails a higher power consumption in CF-mMIMO to further reduce the transmission latency.
In co-located massive MIMO, the MEC servers at the BSs offer huge computational power which is fully exploited by the users, especially those with higher computational demands and poor channel conditions, for which drastically reducing the computational latency is the only way to fulfill the latency requirement.
Changing perspective,~\Figref{fig:fig3}(a) shows the amount of computing resources allocated per MEC server, including APs and CPU. As already mentioned earlier, the ``Small-cell'' resource allocation is conservative and leads to a misuse of the computational resources, which in turn results to an uplink power waste. Notice that the ``Small-cell'' approach attains the minimum energy consumption at the MEC servers---which is proportional to the cube of the computational rates---in line with the objective in~\cite{Mukherjee2020}. The resource allocation for CF-mMIMO via Algorithm~\ref{alg:SCA} leads to excellent uplink power savings with a relative small amount of allocated computational resources per MEC server. On the other hand, the uplink power savings achieved by Algorithm~\ref{alg:cran-scheme}, both for CF-mMIMO and small-cells, can only be obtained by increasing the computational rates, hence the energy consumption, at the MEC servers. As per the co-located setup, the MEC servers basically work at full processing capacity to guarantee the latency requirements.
\begin{figure}[!t]
 	\centering
	\subfloat{\includegraphics[width=.8\columnwidth]{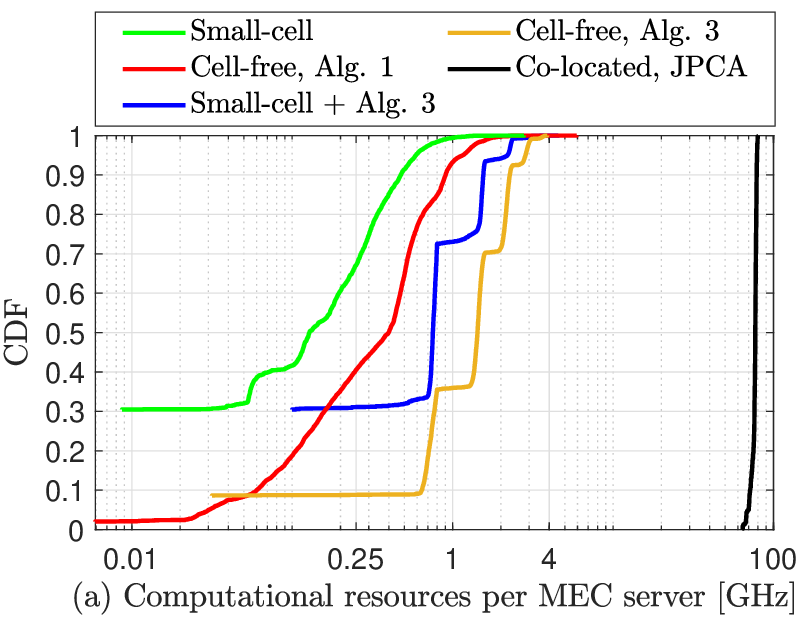}}\\[-0ex]%
    \subfloat{\includegraphics[width=.8\columnwidth]{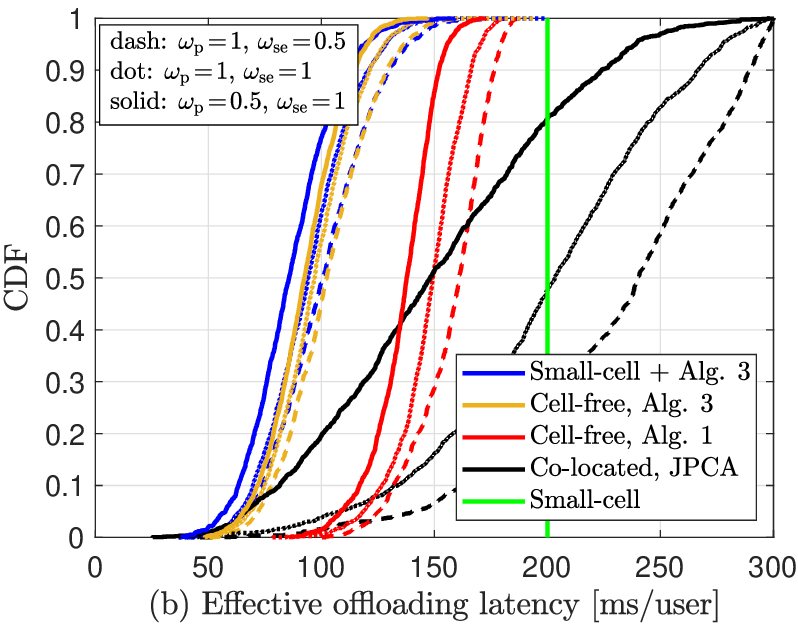}}\vspace*{-1.5ex}
	\caption{(a) Computational resources allocated per MEC server. (b) Effective offloading latency per user.}
	\label{fig:fig3}
\end{figure}\\
\indent The effective latency experienced by the users due to the offloading process is another relevant aspect to measure the effectiveness of the JPCA scheme, and it is shown in~\Figref{fig:fig3}(b). First, we remind that the latency requirements of the cell-free users account for the delay of the data forwarding over the fronthaul network, i.e., step 2 of~\Figref{fig:signalling-diagram}, thus are effectively stricter than those of the co-located and small-cell users.
CF-mMIMO with Algorithm~\ref{alg:SCA} is able to fulfill the latency requirements by a large margin compared to ``Small-cell'' and col-located massive MIMO showing the potentiality to support even stricter requirements. On the other hand, as Algorithm~\ref{alg:cran-scheme} allocates way more computing resources to the users, the computational latency can be remarkably reduced so as to minimize the overall latency experienced by the users. In this regards, the performance of CF-mMIMO and small-cells are almost equivalent when Algorithm~\ref{alg:cran-scheme} is employed. Lastly, notice that the ``Small-cell'' strategy is designed upon satisfying the latency constraint with equality.
The choice of the parameters $\{\op, \ose\}$ clearly affects the effective offloading latency. The latency increases when the uplink power minimization is prioritized over the SE maximization. This suggests that the transmission latency is dominant over the computational latency in this scenario. Importantly, CF-mMIMO combined with Algorithm~\ref{alg:SCA} can simultaneously guarantee significant transmit power saving and low offloading latency, despite the additional delay due to the transmissions over the fronthaul.
\begin{figure}[!t]
	\centering
    \subfloat{\includegraphics[width=.85\columnwidth]{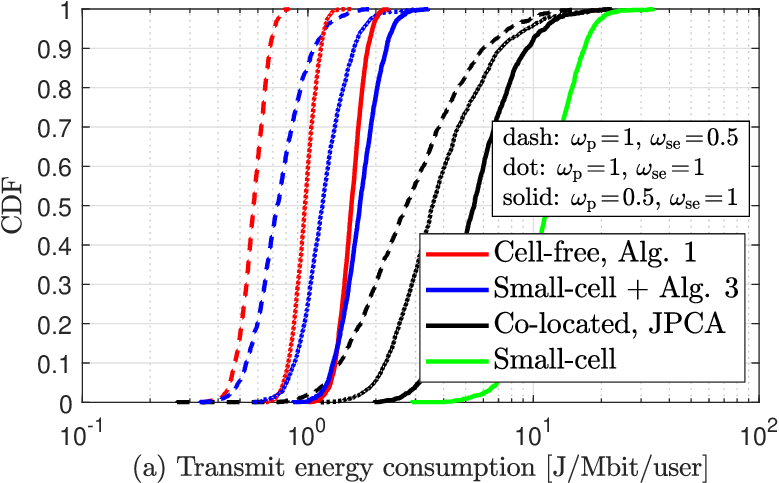}}\\[-0ex]%
    \subfloat{\includegraphics[width=.85\columnwidth]{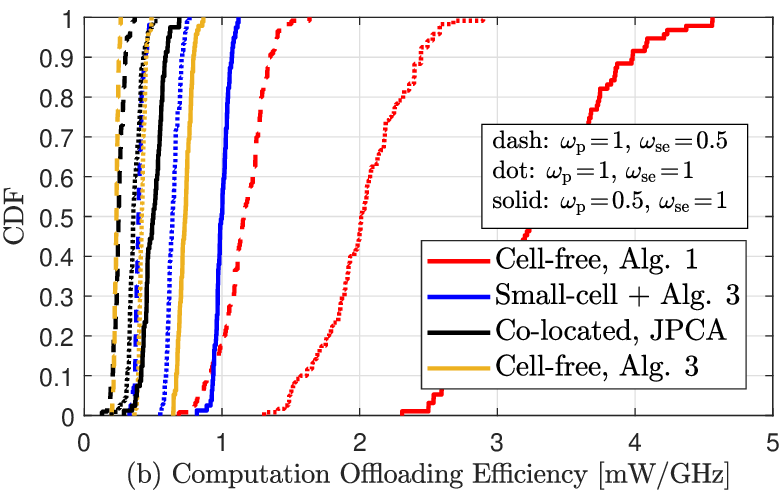}}\vspace*{-1.5ex}
	\caption{(a) Transmit energy consumption in J/Mbit/user. (b) Computation offloading efficiency.}
	\label{fig:fig4}
\end{figure}\\
\indent \Figref{fig:fig4}(a-b) show the transmit energy consumption in J/Mbit/user, given by $\text{E}_k\! =\! p_k/(B \times \text{SE}_k)$, and the computation offloading efficiency (OE), respectively. We define the~OE~as
\begin{equation}
\text{OE} = \frac{\sum\nolimits^K_{k=1} p_k}{\sum\nolimits^K_{k=1}\sum\nolimits^{T_k}_{i=1} f_k(i)} \cdot \frac{\sum\nolimits^K_{k=1} \mathcal{L}_k^{\mathsf{eff}}}{\sum\nolimits^K_{k=1} \mathcal{L}^{\mathsf{req}}_k} \quad \text{[mW/GHz]} \, ,
\end{equation}
where $\mathcal{L}_k^{\mathsf{eff}}$ denotes the effective latency experienced by user $k$ due to the offloading process, i.e., the LHS of the latency constraint in~\eqref{eq:latency-constraint} and~\eqref{prob:P1:cellular:C1} for cell-free and co-located massive MIMO, respectively. While, $\mathcal{L}^{\mathsf{req}}_k$ denotes the latency requirement for user $k$, which is equal to $\mathcal{L}_k$ for CF-mMIMO and small-cells, and equal to $\mathcal{L}^{\mathsf{cell}}_k$ for co-located massive MIMO. 
This metric relates the optimization variables of our interest to each other, and measures the amount of uplink transmit power needed for 1 GHz of computational resources allocated at the MEC servers, also accounting for how much shorter the effective latency is as compared to the requirement. Hence, the larger this metric is, the more efficient the offloading process is.
We observe that cell-free users can save a significant amount of transmit energy with respect to both the co-located and the small-cell users. This confirms the outstanding ability of CF-mMIMO (regardless of the resource allocation algorithm) of simultaneously guarantee low transmission latency and significant transmit energy consumption savings. Not least, fairness among the users is ensured unlike in small-cell and co-located massive MIMO. The energy consumption gap between CF-mMIMO and small-cell is due to the macro-diversity gain provided by the former and increases as we prioritize the power minimization over the SE maximization.
Importantly, the OE attained by CF-mMIMO combined with Algorithm~\ref{alg:SCA} is far superior than any other considered approach. This confirms the nearly-optimal nature of the proposed JPCA strategy over a disjoint radio and computational resource allocation (i.e., Algorithm~\ref{alg:cran-scheme}) as well as over different network setups, namely small-cells and co-located massive MIMO, and despite the stricter latency requirements. Clearly, the OE increases when prioritizing the SE maximization over the uplink power minimization.
\begin{figure}[!t]
	\centering
	\subfloat{\includegraphics[width=.85\columnwidth]{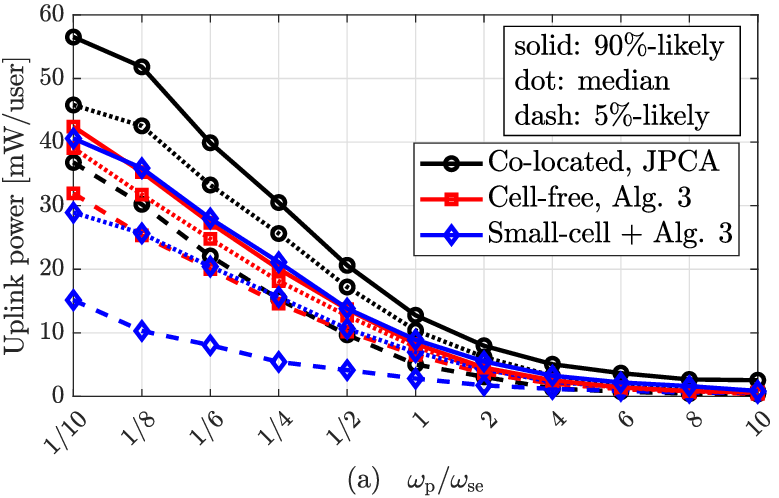}}\\[-0ex]%
	\subfloat{\includegraphics[width=.85\columnwidth]{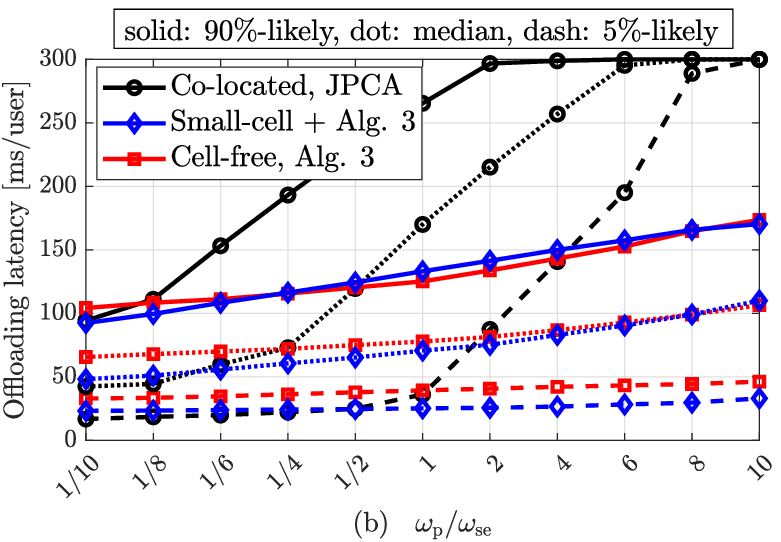}}\vspace*{-1.5ex}
	\caption{(a) Average user transmit power and (b) average offloading latency per user as the ratio $\op/\ose$ varies.}
	\label{fig:fig5}
\end{figure}\\
\indent Finally, we investigate the user transmit power and the effective offloading latency as the ratio $\op/\ose$ varies. By increasing this ratio, the SCA approach solving the optimization with respect to $\bp$ and $\bnu$ prioritizes the minimization of the per-user uplink power, which is clearly shown by the monotonic decreasing behaviour of the curves in~\Figref{fig:fig5}(a).
The uplink power per user achieved in co-located massive MIMO is quite sensitive to the ratio $\op/\ose$ and attains values below 10 mW only for  $\op/\ose\!\geq\! 1$, while the power saving in CF-mMIMO and small-cell via Algorithm~\eqref{alg:cran-scheme} in the region $\op/\ose\!<\!1$ is remarkable. As per the effective offloading latency experienced by the users,~\Figref{fig:fig5}(b) shows that CF-mMIMO and small-cell implementations perform equally well. while the gap with respect to the co-located setup becomes tremendous. Co-located massive MIMO is quite sensitive to the offloading latency as $\op/\ose$ increases, while for CF-mMIMO and small-cell the effective latency increases softly.

\subsection{Pareto Frontier of the proposed MOOP}
\label{subsec:pareto-frontier}
\begin{figure}[!t]
	\centering
	\includegraphics[width=.95\columnwidth]{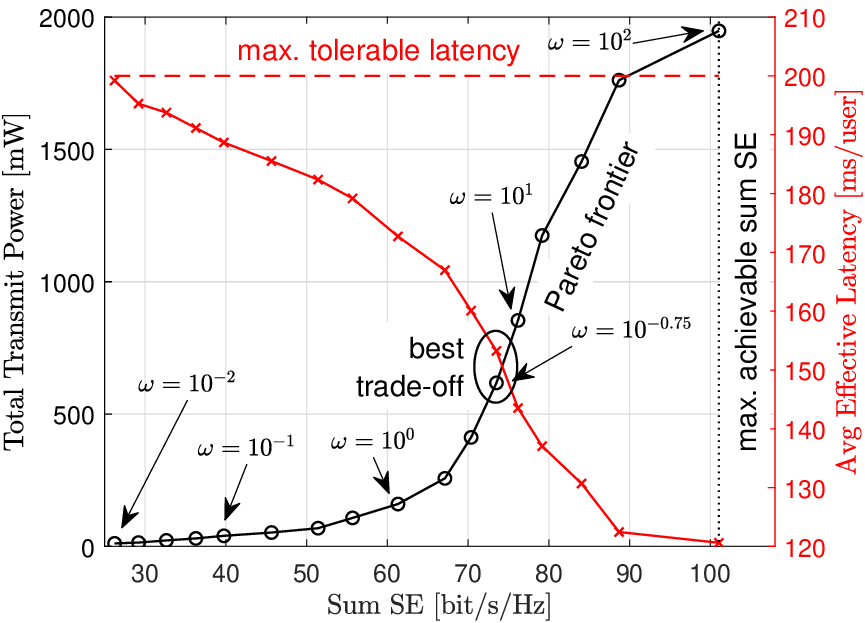}\vspace*{-1.5ex}
	\caption{Sub-optimal Pareto frontier of the JPCA problem in~\eqref{prob:P1} and resulting latency per user. Results obtained by considering the setup in~\Secref{sec:simulation-results}, for one network snapshot and averaging over 200 channel realizations.}
	\label{fig:pareto-frontier}
\end{figure}

As we already mentioned in~\Secref{subsec:JPCA}, there is no unique solution for the SOOP in~\eqref{prob:P1}, but there exists a set of bounded trade-off Pareto optimal solutions, that is a Pareto optimal that enables improvements in some objectives with bounded trade-offs in others. We first reformulate the objective of problem~\eqref{prob:P1} as 
$\vpp \left( \boldsymbol{1}\trans_K \bp \!-\! \boldsymbol{1}\trans_K \bnu {\vpse}/{\vpp} \right) \!=\! \vpp \left( \boldsymbol{1}\trans_K \bp \!-\! \textit{const} \cdot {\omega} \boldsymbol{1}\trans_K \bnu \right),$ 
where ${\omega} \!=\! {\ose}/{\op}$ and $\textit{const} \!=\! {p_{\text{max}}}/{\max\nolimits_k \text{SE}^{(0)}_k}$
are obtained from~\eqref{eq:weights}. Notice that the constant factor $\vpp$ has no effects on the minimization, thus it can be removed from the objective. Finally, a sub-optimal Pareto frontier is obtained by iteratively solving the SOOP according to Algorithm~\ref{alg:SCA} for several values of ${\omega}$ and plotting the corresponding objective values separately, as shown in~\Figref{fig:pareto-frontier} (black curve).
The results in~\Figref{fig:pareto-frontier} are obtained by considering the setup in~\Secref{sec:simulation-results}, for one random realization of APs' and users' locations and averaging over two hundreds random realizations of the small-scale fading.
The Pareto frontier reveals the trade-off between the total transmit power minimization and the sum SE maximization according to the selection of $\omega$ which, in turn, affects the effective latency experienced by the users (red curve). The value of the design parameter $\omega$ that provides the best trade-offs between power saving and latency can be easily identified by inspection from~\Figref{fig:pareto-frontier}.

\subsection{Impact of the AP Selection Strategy}
\label{subsec:AP-selection}
The simulation results shown in this section aim at highlighting the impact of the AP selection strategy on the JPCA scheme in CF-mMIMO. In section~\Secref{subsec:performance-comparison}, we assumed the AP-to-user association described in~\cite[Sec. 5.4]{cellfreebook}, so as to ensure that users served by the same set of the best APs are given orthogonal pilots. Hence, if $\tp=5$, each AP participates to the service of up to 5 users. We refer to this scheme as \textit{dynamic cooperation clustering} (DCC). From an energy efficiency viewpoint such an AP selection strategy results to be costly as many APs, even those bringing negligible contribution to the performance, are involved and active both in the radio communication and computational offloading service of a user.
We next give a qualitative study of the energy consumption at the server side, by considering alternative AP selection strategies. An AP selection strategy establishes a different fraction of APs involved in the communication service. We consider a fixed cooperation clustering (FCC) scheme, wherein each user, upon the associations established by the DCC scheme, is only served by the best (channel-wise) $G$ APs. In addition, we consider the large-scale-fading-based AP selection (LSFBS)~\cite{Ngo2018a}, wherein each user, upon the associations established by the DCC scheme, is only served by the APs that contribute to the 95\% of its channel gain.
\begin{figure}[!t]
	\centering
	\includegraphics[width=\columnwidth]{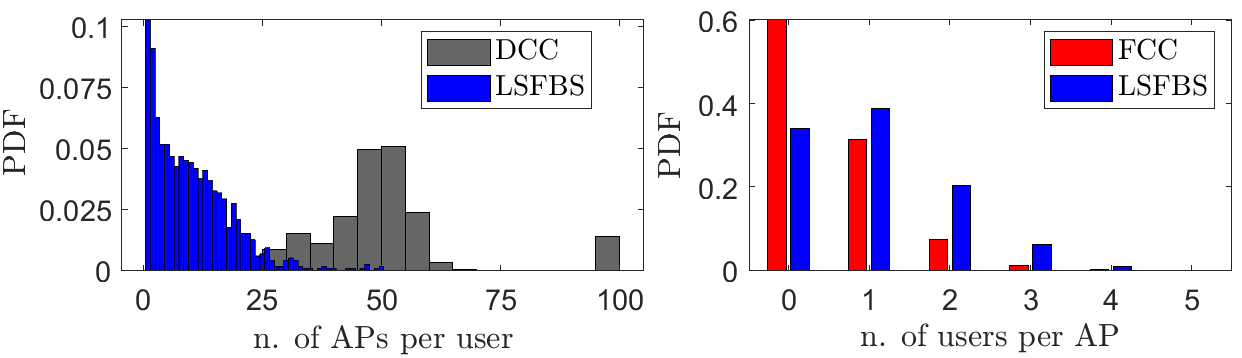}\vspace*{-1.5ex}
	\caption{PDF of the number of (a) APs per user, and (b) users per AP, for different AP selection strategies. $K\!=\!10, \tp\!=\!G\!=\!5$.}
	\label{fig:fig6}
	\vspace*{-5mm}
\end{figure}%
\begin{figure}[!t]
	\centering
	\vspace*{5mm}
	\includegraphics[width=\columnwidth]{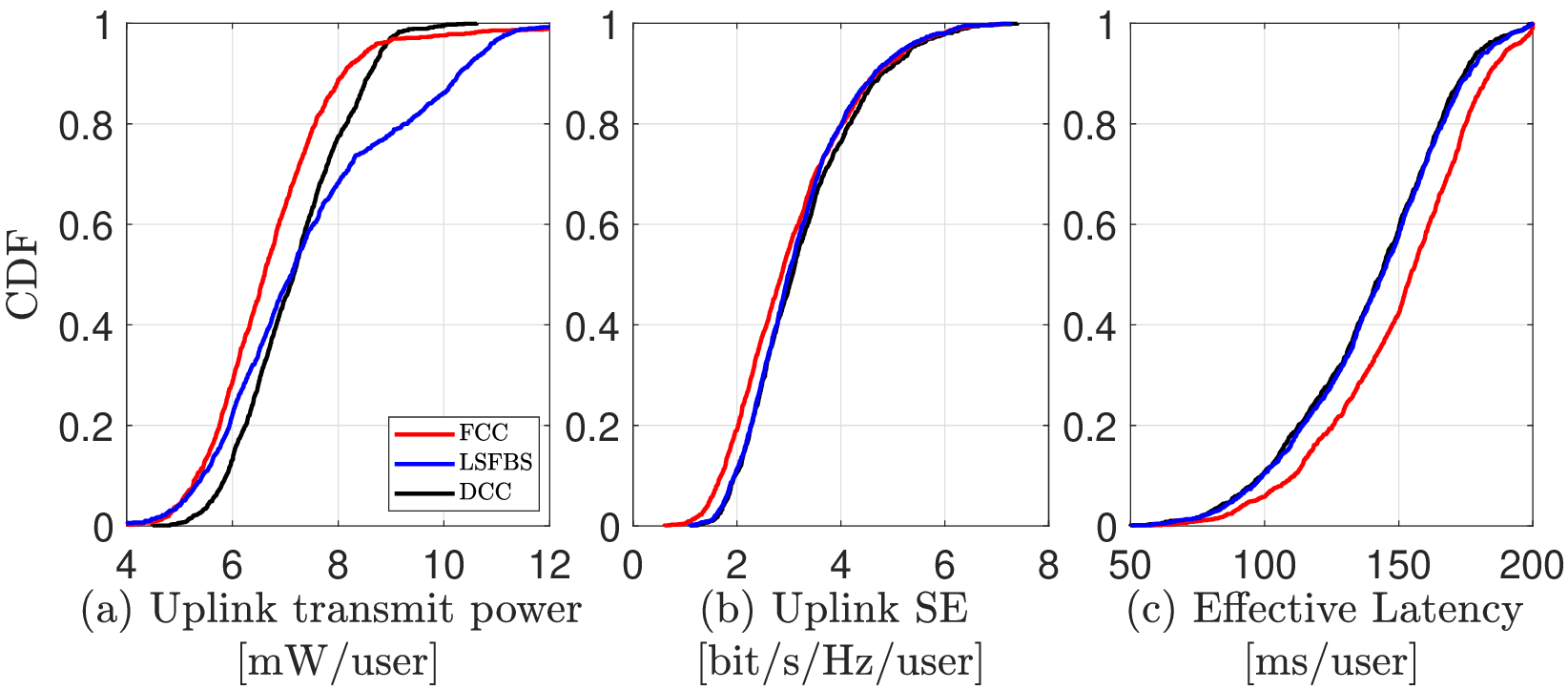}\vspace*{-1ex}
	\caption{CDFs of (a) uplink transmit power per user (b) uplink per-user SE (c) effective offloading latency per user, for different AP selection strategies. $K\!=\!10, \tp\!=\!G\!=\!5$ and $\op=\ose=1$.}
	\label{fig:fig7}
\end{figure}
\Figref{fig:fig6} shows the probability density function (PDF) of the number of APs per user and the number of users per AP, for different AP selection strategies, assuming $K = 10, \tp\!=\!G\!=\!5$. The LSFBS involves a handful of APs per user with high probability, while the DCC scheme selects many APs per user, with a non-negligible probability of selecting all the APs. The FCC scheme always select $G\!=\!5$ APs per user. Changing perspective, 60\% and about 38\% of the APs is off the communication service with FCC and LSFBS, respectively, while the DCC always selects $\tp\!=\!5$ users per AP.
As we can observe in~\Figref{fig:fig7}, selecting a fixed number of APs per user is not convenient, as each user needs a tailored number of cooperating APs for achieving the cell-free experience. Hence, the FCC AP selection strategy achieves an uplink SE slightly lower than DCC and LSFBS which, in turn, results in a longer effective latency. While, DCC and LSFBS strategies provide equivalent per-user SE and effective latency, although with different levels of uplink transmit power per user. To counterbalance the lack of macro-diversity gain when selecting a very few APs, LSFBS requires the users to use more transmit power than DCC, SE being equal. Conversely, when the DCC selects too many APs, a user needs to use higher uplink powers to guarantee a good receive combining at the furthest APs.
Lastly, the AP selection strategy only concerns the communication service, thus it has no impact on the computing resource allocation, not shown herein for brevity.  

\subsection{JPCA: A Closer Outlook} \label{subsec:closer-outlook}
\begin{figure*}[!t]
\centering
\includegraphics[width=\linewidth]{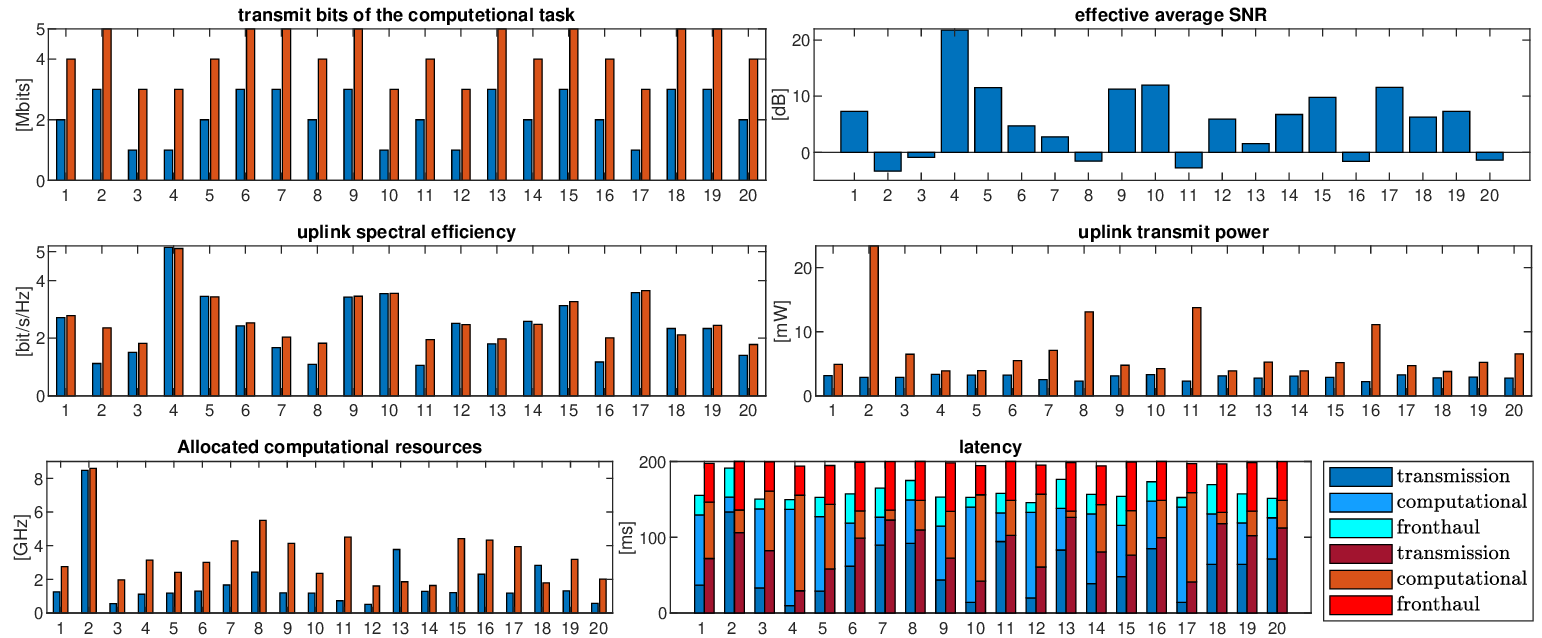}\vspace*{-1ex}
\caption{Simulation results for one network snapshot, averaged over 200 channel realizations. $x$-axes report the user index. Blueish bars refer to the cell-free setup described in Section~\ref{sec:simulation-results}, but with $\{b_k\} \!\sim\! \mathcal{U}(1, 3)$. Reddish bars refer to a cell-free setup with $\{b_k\} \!\sim\! \mathcal{U}(3, 5)$ Mbits. Black bars refer to both the setups.}
\label{fig:fig9}
\end{figure*}

In this section, we explore more in detail the effectiveness of the proposed JPCA, evaluating the interplay between uplink power, SE, allocated computational resources, and effective offloading latency. In \Figref{fig:fig9}, we present the simulation results of one network snapshot, where the final values are averaged over 200 channel realizations.
For all the users, whose indices appear on the $x$-axes, we report: computational task size $b_k$ in Mbits; effective average SNR computed as $\sum_{l \in \mathcal{M}_k} \beta_{lk}/\sigma^2$ and converted to dB; uplink SE in bit/s/Hz; uplink transmit power in mW; the allocated computational resources per user, namely $\{\sum^{T_k}_{i=1}f_k(i)\}$ in GHz; and the effective offloading latency consisting of transmission, computational and fronthaul latency. We consider two simulation setups: blueish bars refer to the cell-free setup described in Section~\ref{sec:simulation-results} but with $\{b_k\} \sim \mathcal{U}(1, 3)$, which we call ``loose'' setup for brevity; reddish bars refer to a ``strict'' cell-free setup with higher user computational demands, that is $\{b_k\} \sim \mathcal{U}(3, 5)$ Mbits. The latter is of particular interest because highlights how the JPCA operates under stricter constraints. The propagation scenario is in common to both the setups, as we fixed the simulation seed in order to obtain the same channel conditions (black bars).
By inspecting \Figref{fig:fig9}, we observe that user 2 is in the adverse conditions of poor SNR and high computational demand. The JPCA naturally needs to allocate more power and computational resources to this user than others in order to reduce its transmission latency (by increasing its SE) and computational latency. As a comparison, user 9 has the same computational demand but better SNR, hence its latency requirements can be more easily fulfilled by solely reducing its transmission latency through allocating slightly more uplink power.
In the ``strict'' setup (reddish bars) the fronthaul latency is longer as it is proportional to the user's task size, and the overall effective latency almost equals the user's requirements of 200 ms.
For the ``loose'' setup (blueish bars) we obtained different but equally interesting results. The fronthaul latency is less pronounced due to the lower user's computational demands. Importantly, we observe a more uniform allocation of the uplink power over the users as compared to the ``strict'' setup case.
The computational latency is dominant over the transmission latency for those users experiencing good channel conditions as high SEs can be achieved with small amount of transmit powers. Conversely, the transmission latency is dominant for those users experiencing bad channel quality.
Interestingly, we observe that the effective offloading latency is far below the latency requirement of 200 ms, which results from the choice of achieving a fair balance between transmit power saving and offloading latency by setting $\op=\ose=1$.

\section{Conclusion}\label{sec:conclusion}
The problem of jointly allocating the uplink powers and network computational resources subject to latency constraints in a MEC-enabled CF-mMIMO system was considered in this paper, with the aim of minimizing the total transmit power and simultaneously maximizing the uplink sum SE, and thereby providing an excellent trade-off between user power consumption and effective offloading latency. 
For efficiently solving such a non-convex problem, a framework based on alternating optimization and successive convex approximation along with an alternative low-complexity heuristic approach were proposed. A detailed performance comparison between the proposed MEC-enabled CF-mMIMO architecture, its co-located and small-cell counterparts was also provided. Simulation results revealed that CF-mMIMO provides far superior computation offloading efficiency than other network architectures, and constitutes a promising candidate to suitably and flexibly support MEC applications. The proposed joint resource allocation strategy is effective in simultaneously guaranteeing to the users low offloading latency, fairness and significant transmit power saving by distributing the computational workload over multiple MEC servers.
Devising a low-complexity JPCA algorithm based on learning~\cite{Guo2022,Tang2023} and/or non-convex optimization (e.g., \textit{differential evolution}~\cite{Yu2022}) is an appealing research direction for future works, as well as extending this study to a partial computational offloading model.

\bibliographystyle{IEEEtran}
\bibliography{IEEEabrv,refs-abbr}

\enlargethispage{-4in}
\begin{IEEEbiography}[{\includegraphics[width=1in,height=1.25in,clip,keepaspectratio]{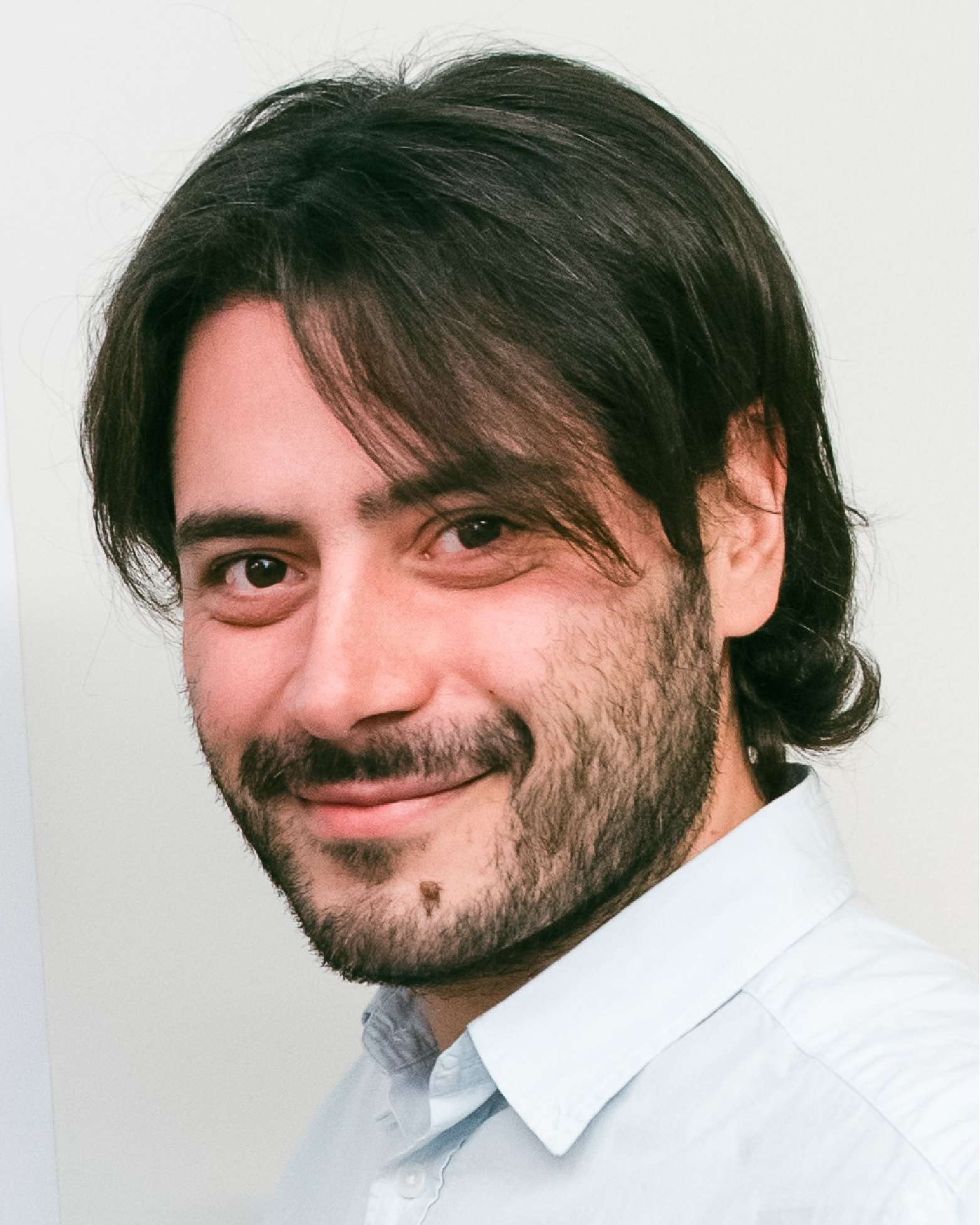}}]{Giovanni Interdonato}
(Member, IEEE) received the M.Sc. degree in computer and telecommunication systems engineering from the Mediterranea University of Reggio Calabria, Italy, in 2015, 
and the Ph.D. degree in electrical engineering with a specialization in communication systems from Link\"{o}ping University, Sweden, in 2020. 
From October 2015 to October 2018, he was a researcher at the Radio Network Department of Ericsson Research, Link\"{o}ping, and a Marie Sklodowska-Curie research fellow of the EU-H2020 ITN project ``5Gwireless''.

Dr. Interdonato is currently an assistant professor at the Department of Electrical and Information Engineering (DIEI), University of Cassino and Southern Lazio, Italy.
His main research interests lie in the field of wireless communications and signal processing, with focus on beyond-5G physical layer technologies, radio resource management and communication protocols. 
He is the co-inventor of about twenty granted patent applications on massive MIMO and cell-free massive MIMO systems.
Dr. Interdonato serves as an associate editor for \textsc{IEEE Communications Letters} and \textsc{IEEE Open Journal of the Communications Society}. 

He has been awarded \textsc{IEEE Transactions on Communications} Exemplary Reviewer for 2021 and \textsc{IEEE Communications Letters} Exemplary Editor for 2022, and he was a recipient of a research grant from the Ericsson Research Foundation in 2019.    
\end{IEEEbiography}

\vspace*{-8mm}
\begin{IEEEbiography}[{\includegraphics[width=1in,height=1.25in,clip,keepaspectratio]{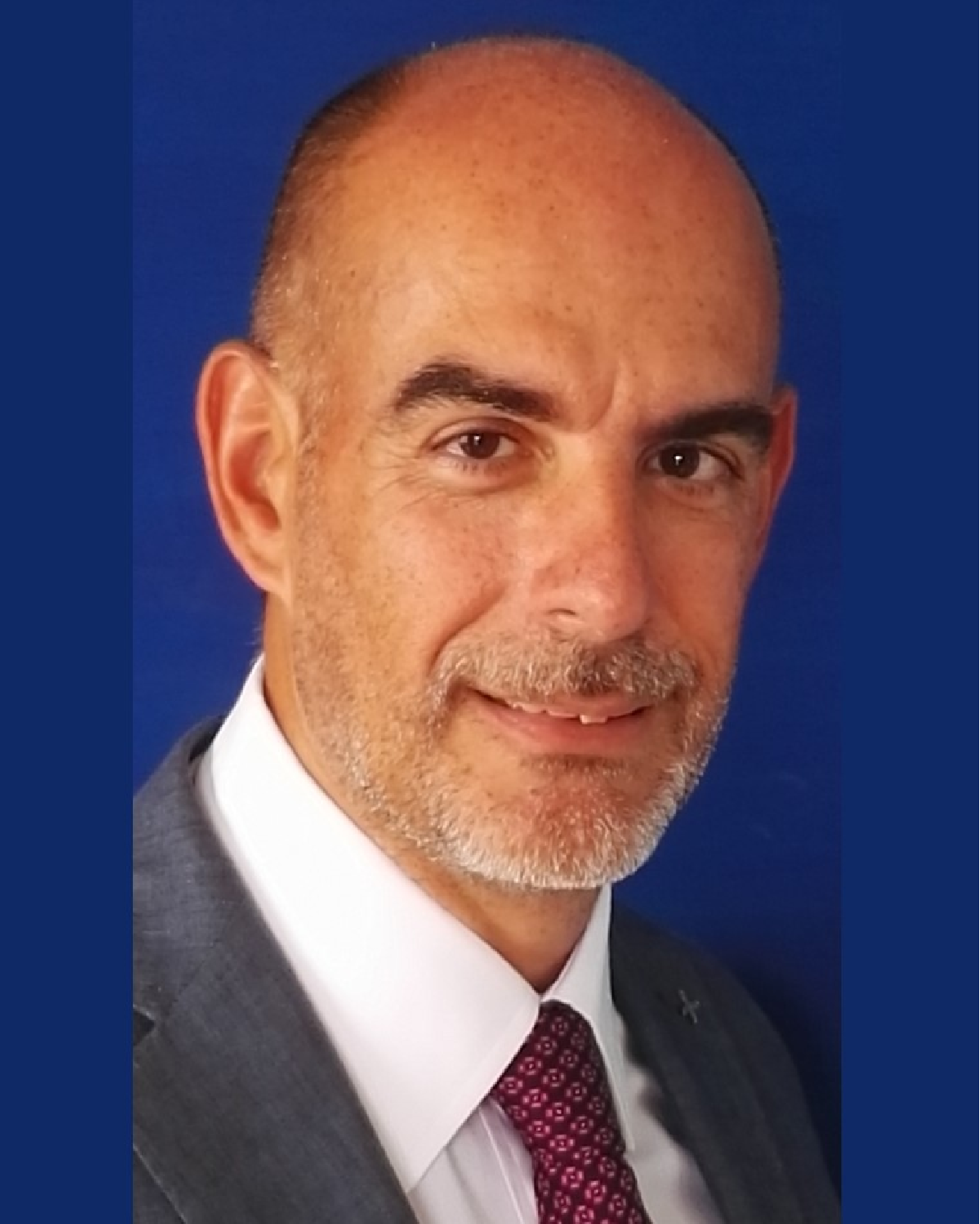}}]{Stefano Buzzi} (Senior Member, IEEE) joined the University of Cassino and Lazio Meridionale, Italy in 2000, first as an Assistant Professor, then as an Associate Professor (since 2002)  and, finally, since 2018, as a Full Professor. 
He received the M.Sc. degree (summa cum laude)  in Electronic Engineering in 1994, and the Ph.D. degree in Electrical and Computer Engineering in 1999, both from the University of Naples ``Federico II''.  He has had short-term research appointments at Princeton University, Princeton (NJ), USA in 1999, 2000, 2001 and 2006. He is a former Associate Editor of the \textsc{IEEE Signal Processing Letters} and of the \textsc{IEEE Communications Letters}, has been the guest editor of four \textsc{IEEE JSAC} special issues, and from 2014  to 2020 he has been an Editor for the \textsc{IEEE Transactions on Wireless Communications}. 

Currently, Prof. Buzzi is Associate Editor for the \textsc{IEEE Transactions on Communications}. He also serves regularly as TPC member of several international conferences.
Dr. Buzzi's research interests are in the broad field of communications and signal processing, with emphasis on wireless communications and beyond-5G systems.  He is currently the General Coordinator of the EU-funded Innovative Training Network project METAWIRELESS, on the application of metasurfaces to wireless communications, and of the EU-funded Doctoral Network ISLANDS, on Integrated Sensing and Communications for the Vehicular Environment. He has co-authored about 180 technical peer-reviewed journal and conference papers, and, among these, the highly cited paper ``What will 5G be?'', \textsc{IEEE JSAC}, June 2014.
\end{IEEEbiography}
\vfill

\end{document}

%% file: report-macros.tex
\usepackage{amsmath,amssymb}
\usepackage{mathtools} 
\usepackage{xspace}

\newcommand{\eqrefs}[2]{equations~(\ref{#1})-(\ref{#2})}
\newcommand{\Secref}[1]{Section~\ref{#1}}
\newcommand{\Figref}[1]{Fig.~\ref{#1}}

\newcommand{\herm}{^\text{H}}

\newcommand{\trans}{^\text{T}}

\newcommand{\bZ}{\mathbf{Z}}

\newcommand{\bh}{\mathbf{h}}

\newcommand{\bG}{\mathbf{G}}

\newcommand{\bY}{\mathbf{Y}}
\newcommand{\by}{\mathbf{y}}
\newcommand{\bI}{\mathbf{I}}

\newcommand{\bR}{\mathbf{R}}

\newcommand{\bC}{\mathbf{C}}
\newcommand{\bD}{\mathbf{D}}

\newcommand{\bc}{\mathbf{c}}

\newcommand{\bv}{\mathbf{v}}

\newcommand{\bu}{\mathbf{u}}
\newcommand{\bn}{\mathbf{n}}

\newcommand{\bp}{\mathbf{p}}

\newcommand{\bOmega}{\boldsymbol{\Omega}}

\newcommand{\bvphi}{\boldsymbol{\varphi}}

\newcommand{\bPsi}{\boldsymbol{\Psi}}

\newcommand{\bzero}{\boldsymbol{0}}
\newcommand{\bnu}{\boldsymbol{\nu}}

\newcommand{\CN}{\mathcal{CN}}
\newcommand{\tc}{\tau_{\textsc{c}}}

\newcommand{\boundellipse}[3]
{(#1) ellipse [x radius=#2,y radius=#3]
}

\DeclareMathOperator{\tr}{tr}

\DeclareMathOperator*{\argmax}{arg\,max}

\newcommand{\EX}[1]{\mathsf{E}\left\{{#1}\right\}}

\newcommand{\C}{\mathbb{C}}

\newcommand{\op}{\omega_{\mathrm{p}}}
\newcommand{\ose}{\omega_{\mathrm{se}}}
\newcommand{\vpp}{\varpi_{\mathrm{p}}}
\newcommand{\vpse}{\varpi_{\mathrm{se}}}

\newcommand{\tauc}{\tau_\mathrm{c}}

\newcommand{\norm}[1]{{ \left\Vert #1 \right\Vert }}

\newcommand{\tp}{\tau_{\mathrm{p}}}
\newcommand{\td}{\tau_{\mathrm{d}}}
\newcommand{\tu}{\tau_{\mathrm{u}}}

\makeatletter
\def\@setsize#1#2#3#4{
    \@nomath#1
    \let\@currsize#1
    \baselineskip #2
    \baselineskip \baselinestretch\baselineskip
    \parskip \baselinestretch\parskip
    \setbox\strutbox \hbox{
        \vrule height.7\baselineskip
            depth.3\baselineskip
            width\z@}
    \skip\footins \baselinestretch\skip\footins
    \normalbaselineskip\baselineskip#3#4}
\makeatother

\makeatletter
\newcommand{\setstretch}[1]{
    \def\baselinestretch{#1}%
    \@currsize
    }
\makeatother

%% file: signalling-diagram.tex
\tikzset{every picture/.style={line width=0.75pt}} 

\begin{tikzpicture}[x=0.75pt,y=0.75pt,yscale=-1,xscale=1]

\draw  [color={rgb, 255:red, 0; green, 0; blue, 0 }  ,draw opacity=1 ][fill={rgb, 255:red, 255; green, 255; blue, 255 }  ,fill opacity=1 ] (186,57.5) .. controls (186,56.4) and (186.9,55.5) .. (188,55.5) -- (224,55.5) .. controls (225.1,55.5) and (226,56.4) .. (226,57.5) -- (226,63.5) .. controls (226,64.6) and (225.1,65.5) .. (224,65.5) -- (188,65.5) .. controls (186.9,65.5) and (186,64.6) .. (186,63.5) -- cycle ;
\draw  [color={rgb, 255:red, 0; green, 0; blue, 0 }  ,draw opacity=1 ][fill={rgb, 255:red, 255; green, 255; blue, 255 }  ,fill opacity=1 ] (186,67.5) .. controls (186,66.4) and (186.9,65.5) .. (188,65.5) -- (224,65.5) .. controls (225.1,65.5) and (226,66.4) .. (226,67.5) -- (226,73.5) .. controls (226,74.6) and (225.1,75.5) .. (224,75.5) -- (188,75.5) .. controls (186.9,75.5) and (186,74.6) .. (186,73.5) -- cycle ;
\draw  [color={rgb, 255:red, 0; green, 0; blue, 0 }  ,draw opacity=1 ][fill={rgb, 255:red, 255; green, 255; blue, 255 }  ,fill opacity=1 ] (186,47.5) .. controls (186,46.4) and (186.9,45.5) .. (188,45.5) -- (224,45.5) .. controls (225.1,45.5) and (226,46.4) .. (226,47.5) -- (226,53.5) .. controls (226,54.6) and (225.1,55.5) .. (224,55.5) -- (188,55.5) .. controls (186.9,55.5) and (186,54.6) .. (186,53.5) -- cycle ;
\draw  [color={rgb, 255:red, 155; green, 155; blue, 155 }  ,draw opacity=1 ] (315.3,51.32) .. controls (315.3,50.72) and (315.79,50.23) .. (316.39,50.23) -- (335.21,50.23) .. controls (335.81,50.23) and (336.3,50.72) .. (336.3,51.32) -- (336.3,55.7) .. controls (336.3,55.7) and (336.3,55.7) .. (336.3,55.7) -- (315.3,55.7) .. controls (315.3,55.7) and (315.3,55.7) .. (315.3,55.7) -- cycle ;
\draw  [color={rgb, 255:red, 155; green, 155; blue, 155 }  ,draw opacity=1 ] (317.31,52.85) .. controls (317.31,52.2) and (317.84,51.68) .. (318.48,51.68) .. controls (319.13,51.68) and (319.66,52.2) .. (319.66,52.85) .. controls (319.66,53.5) and (319.13,54.02) .. (318.48,54.02) .. controls (317.84,54.02) and (317.31,53.5) .. (317.31,52.85) -- cycle ;
\draw  [color={rgb, 255:red, 155; green, 155; blue, 155 }  ,draw opacity=1 ] (321.11,52.85) .. controls (321.11,52.2) and (321.63,51.68) .. (322.28,51.68) .. controls (322.93,51.68) and (323.45,52.2) .. (323.45,52.85) .. controls (323.45,53.5) and (322.93,54.02) .. (322.28,54.02) .. controls (321.63,54.02) and (321.11,53.5) .. (321.11,52.85) -- cycle ;
\draw [color={rgb, 255:red, 155; green, 155; blue, 155 }  ,draw opacity=1 ]   (325.8,52.96) -- (333.51,52.96) ;
\draw [color={rgb, 255:red, 155; green, 155; blue, 155 }  ,draw opacity=1 ]   (318.87,39.5) -- (318.99,50.34) ;
\draw [shift={(318.87,39.5)}, rotate = 89.41] [color={rgb, 255:red, 155; green, 155; blue, 155 }  ,draw opacity=1 ][fill={rgb, 255:red, 155; green, 155; blue, 155 }  ,fill opacity=1 ][line width=0.75]      (0, 0) circle [x radius= 1.34, y radius= 1.34]   ;
\draw [color={rgb, 255:red, 155; green, 155; blue, 155 }  ,draw opacity=1 ]   (332.73,39.5) -- (332.84,50.34) ;
\draw [shift={(332.73,39.5)}, rotate = 89.41] [color={rgb, 255:red, 155; green, 155; blue, 155 }  ,draw opacity=1 ][fill={rgb, 255:red, 155; green, 155; blue, 155 }  ,fill opacity=1 ][line width=0.75]      (0, 0) circle [x radius= 1.34, y radius= 1.34]   ;

\draw  [color={rgb, 255:red, 155; green, 155; blue, 155 }  ,draw opacity=1 ] (335.3,32.1) .. controls (335.3,31.55) and (335.75,31.1) .. (336.3,31.1) -- (353.49,31.1) .. controls (354.04,31.1) and (354.49,31.55) .. (354.49,32.1) -- (354.49,36.1) .. controls (354.49,36.1) and (354.49,36.1) .. (354.49,36.1) -- (335.3,36.1) .. controls (335.3,36.1) and (335.3,36.1) .. (335.3,36.1) -- cycle ;
\draw  [color={rgb, 255:red, 155; green, 155; blue, 155 }  ,draw opacity=1 ] (337.14,33.5) .. controls (337.14,32.91) and (337.62,32.43) .. (338.21,32.43) .. controls (338.8,32.43) and (339.28,32.91) .. (339.28,33.5) .. controls (339.28,34.09) and (338.8,34.57) .. (338.21,34.57) .. controls (337.62,34.57) and (337.14,34.09) .. (337.14,33.5) -- cycle ;
\draw  [color={rgb, 255:red, 155; green, 155; blue, 155 }  ,draw opacity=1 ] (340.61,33.5) .. controls (340.61,32.91) and (341.09,32.43) .. (341.68,32.43) .. controls (342.27,32.43) and (342.75,32.91) .. (342.75,33.5) .. controls (342.75,34.09) and (342.27,34.57) .. (341.68,34.57) .. controls (341.09,34.57) and (340.61,34.09) .. (340.61,33.5) -- cycle ;
\draw [color={rgb, 255:red, 155; green, 155; blue, 155 }  ,draw opacity=1 ]   (344.89,33.6) -- (351.94,33.6) ;
\draw [color={rgb, 255:red, 155; green, 155; blue, 155 }  ,draw opacity=1 ]   (338.57,21.3) -- (338.67,31.2) ;
\draw [shift={(338.57,21.3)}, rotate = 89.41] [color={rgb, 255:red, 155; green, 155; blue, 155 }  ,draw opacity=1 ][fill={rgb, 255:red, 155; green, 155; blue, 155 }  ,fill opacity=1 ][line width=0.75]      (0, 0) circle [x radius= 1.34, y radius= 1.34]   ;
\draw [color={rgb, 255:red, 155; green, 155; blue, 155 }  ,draw opacity=1 ]   (351.22,21.3) -- (351.32,31.2) ;
\draw [shift={(351.22,21.3)}, rotate = 89.41] [color={rgb, 255:red, 155; green, 155; blue, 155 }  ,draw opacity=1 ][fill={rgb, 255:red, 155; green, 155; blue, 155 }  ,fill opacity=1 ][line width=0.75]      (0, 0) circle [x radius= 1.34, y radius= 1.34]   ;

\draw  [color={rgb, 255:red, 155; green, 155; blue, 155 }  ,draw opacity=1 ] (354.1,50.04) .. controls (354.1,49.42) and (354.6,48.92) .. (355.22,48.92) -- (374.38,48.92) .. controls (375,48.92) and (375.5,49.42) .. (375.5,50.04) -- (375.5,54.5) .. controls (375.5,54.5) and (375.5,54.5) .. (375.5,54.5) -- (354.1,54.5) .. controls (354.1,54.5) and (354.1,54.5) .. (354.1,54.5) -- cycle ;
\draw  [color={rgb, 255:red, 155; green, 155; blue, 155 }  ,draw opacity=1 ] (356.15,51.6) .. controls (356.15,50.94) and (356.68,50.4) .. (357.34,50.4) .. controls (358,50.4) and (358.54,50.94) .. (358.54,51.6) .. controls (358.54,52.26) and (358,52.79) .. (357.34,52.79) .. controls (356.68,52.79) and (356.15,52.26) .. (356.15,51.6) -- cycle ;
\draw  [color={rgb, 255:red, 155; green, 155; blue, 155 }  ,draw opacity=1 ] (360.02,51.6) .. controls (360.02,50.94) and (360.55,50.4) .. (361.21,50.4) .. controls (361.87,50.4) and (362.41,50.94) .. (362.41,51.6) .. controls (362.41,52.26) and (361.87,52.79) .. (361.21,52.79) .. controls (360.55,52.79) and (360.02,52.26) .. (360.02,51.6) -- cycle ;
\draw [color={rgb, 255:red, 155; green, 155; blue, 155 }  ,draw opacity=1 ]   (364.8,51.71) -- (372.65,51.71) ;
\draw [color={rgb, 255:red, 155; green, 155; blue, 155 }  ,draw opacity=1 ]   (357.74,37.99) -- (357.86,49.04) ;
\draw [shift={(357.74,37.99)}, rotate = 89.41] [color={rgb, 255:red, 155; green, 155; blue, 155 }  ,draw opacity=1 ][fill={rgb, 255:red, 155; green, 155; blue, 155 }  ,fill opacity=1 ][line width=0.75]      (0, 0) circle [x radius= 1.34, y radius= 1.34]   ;
\draw [color={rgb, 255:red, 155; green, 155; blue, 155 }  ,draw opacity=1 ]   (371.86,37.99) -- (371.97,49.04) ;
\draw [shift={(371.86,37.99)}, rotate = 89.41] [color={rgb, 255:red, 155; green, 155; blue, 155 }  ,draw opacity=1 ][fill={rgb, 255:red, 155; green, 155; blue, 155 }  ,fill opacity=1 ][line width=0.75]      (0, 0) circle [x radius= 1.34, y radius= 1.34]   ;

\draw    (489.8,100) -- (489.8,259.01) ;
\draw   (487.37,48.6) .. controls (488.84,48.6) and (490.03,49.79) .. (490.03,51.27) -- (490.03,67.93) .. controls (490.03,69.41) and (488.84,70.6) .. (487.37,70.6) -- (479.37,70.6) .. controls (477.89,70.6) and (476.7,69.41) .. (476.7,67.93) -- (476.7,51.27) .. controls (476.7,49.79) and (477.89,48.6) .. (479.37,48.6) -- cycle ;
\draw   (485.23,66.87) .. controls (485.23,65.8) and (484.37,64.93) .. (483.3,64.93) .. controls (482.23,64.93) and (481.37,65.8) .. (481.37,66.87) .. controls (481.37,67.93) and (482.23,68.8) .. (483.3,68.8) .. controls (484.37,68.8) and (485.23,67.93) .. (485.23,66.87) -- cycle ;

\draw  [dash pattern={on 0.75pt off 0.75pt}]  (375.9,61.7) .. controls (376.47,59.3) and (377.92,58.36) .. (380.24,58.87) .. controls (382.67,59.43) and (384.13,58.66) .. (384.63,56.57) .. controls (385.55,54.4) and (387.13,53.84) .. (389.37,54.89) .. controls (391.15,56.24) and (392.78,56.04) .. (394.26,54.28) .. controls (396.11,52.77) and (397.75,53.05) .. (399.18,55.1) .. controls (400.07,57.25) and (401.55,57.92) .. (403.62,57.13) .. controls (406.07,56.7) and (407.48,57.64) .. (407.87,59.95) .. controls (408.23,62.26) and (409.59,63.25) .. (411.95,62.9) .. controls (414.31,62.5) and (415.67,63.4) .. (416.04,65.61) .. controls (416.87,67.99) and (418.41,68.79) .. (420.64,68.01) .. controls (422.75,67.06) and (424.37,67.62) .. (425.5,69.71) .. controls (426.68,71.68) and (428.22,71.97) .. (430.12,70.57) .. controls (431.88,69.04) and (433.62,69.12) .. (435.34,70.8) .. controls (436.99,72.39) and (438.55,72.27) .. (440.03,70.45) .. controls (441.42,68.56) and (443.15,68.25) .. (445.2,69.54) .. controls (447.49,70.7) and (449.14,70.27) .. (450.17,68.25) .. controls (451.16,66.19) and (452.7,65.68) .. (454.79,66.72) .. controls (457.02,67.67) and (458.67,67.03) .. (459.75,64.8) .. controls (460.24,62.78) and (461.71,62.13) .. (464.16,62.86) -- (466,62) ;
\draw    (227,66.9) .. controls (313.5,22) and (295.24,91.35) .. (329.5,64.5) ;
\draw   (332.7,62.06) .. controls (332.7,60.98) and (333.58,60.1) .. (334.66,60.1) -- (368.34,60.1) .. controls (369.42,60.1) and (370.3,60.98) .. (370.3,62.06) -- (370.3,69.9) .. controls (370.3,69.9) and (370.3,69.9) .. (370.3,69.9) -- (332.7,69.9) .. controls (332.7,69.9) and (332.7,69.9) .. (332.7,69.9) -- cycle ;
\draw   (336.3,64.8) .. controls (336.3,63.64) and (337.24,62.7) .. (338.4,62.7) .. controls (339.56,62.7) and (340.5,63.64) .. (340.5,64.8) .. controls (340.5,65.96) and (339.56,66.9) .. (338.4,66.9) .. controls (337.24,66.9) and (336.3,65.96) .. (336.3,64.8) -- cycle ;
\draw   (343.1,64.8) .. controls (343.1,63.64) and (344.04,62.7) .. (345.2,62.7) .. controls (346.36,62.7) and (347.3,63.64) .. (347.3,64.8) .. controls (347.3,65.96) and (346.36,66.9) .. (345.2,66.9) .. controls (344.04,66.9) and (343.1,65.96) .. (343.1,64.8) -- cycle ;
\draw    (351.5,65) -- (365.3,65) ;
\draw    (339.1,40.9) -- (339.3,60.3) ;
\draw [shift={(339.1,40.9)}, rotate = 89.41] [color={rgb, 255:red, 0; green, 0; blue, 0 }  ][fill={rgb, 255:red, 0; green, 0; blue, 0 }  ][line width=0.75]      (0, 0) circle [x radius= 1.34, y radius= 1.34]   ;
\draw    (363.9,40.9) -- (364.1,60.3) ;
\draw [shift={(363.9,40.9)}, rotate = 89.41] [color={rgb, 255:red, 0; green, 0; blue, 0 }  ][fill={rgb, 255:red, 0; green, 0; blue, 0 }  ][line width=0.75]      (0, 0) circle [x radius= 1.34, y radius= 1.34]   ;

\draw    (353.78,114.5) -- (489.8,100) ;
\draw [shift={(350.8,114.82)}, rotate = 353.91] [fill={rgb, 255:red, 0; green, 0; blue, 0 }  ][line width=0.08]  [draw opacity=0] (8.93,-4.29) -- (0,0) -- (8.93,4.29) -- cycle    ;
\draw    (203.98,151.69) -- (351,136.01) ;
\draw [shift={(201,152.01)}, rotate = 353.91] [fill={rgb, 255:red, 0; green, 0; blue, 0 }  ][line width=0.08]  [draw opacity=0] (8.93,-4.29) -- (0,0) -- (8.93,4.29) -- cycle    ;
\draw   (197.2,101.1) .. controls (192.53,101.08) and (190.19,103.4) .. (190.17,108.07) -- (190.14,117.08) .. controls (190.11,123.75) and (187.77,127.07) .. (183.1,127.06) .. controls (187.77,127.07) and (190.09,130.41) .. (190.06,137.08)(190.07,134.08) -- (190.02,144.98) .. controls (190.01,149.65) and (192.33,151.99) .. (197,152.01) ;
\draw [color={rgb, 255:red, 0; green, 0; blue, 0 }  ,draw opacity=1 ]   (206,48.5) -- (219.8,48.5) ;
\draw [color={rgb, 255:red, 0; green, 0; blue, 0 }  ,draw opacity=1 ]   (206,50.5) -- (219.8,50.5) ;
\draw [color={rgb, 255:red, 0; green, 0; blue, 0 }  ,draw opacity=1 ]   (206,52.5) -- (219.8,52.5) ;
\draw [color={rgb, 255:red, 0; green, 0; blue, 0 }  ,draw opacity=1 ]   (206.4,58.5) -- (220.2,58.5) ;
\draw [color={rgb, 255:red, 0; green, 0; blue, 0 }  ,draw opacity=1 ]   (206.4,60.5) -- (220.2,60.5) ;
\draw [color={rgb, 255:red, 0; green, 0; blue, 0 }  ,draw opacity=1 ]   (206.4,62.5) -- (220.2,62.5) ;
\draw [color={rgb, 255:red, 0; green, 0; blue, 0 }  ,draw opacity=1 ]   (206.4,68.5) -- (220.2,68.5) ;
\draw [color={rgb, 255:red, 0; green, 0; blue, 0 }  ,draw opacity=1 ]   (206.4,70.5) -- (220.2,70.5) ;
\draw [color={rgb, 255:red, 0; green, 0; blue, 0 }  ,draw opacity=1 ]   (206.4,72.5) -- (220.2,72.5) ;
\draw  [color={rgb, 255:red, 0; green, 0; blue, 0 }  ,draw opacity=1 ] (189.6,50.4) .. controls (189.6,49.24) and (190.54,48.3) .. (191.7,48.3) .. controls (192.86,48.3) and (193.8,49.24) .. (193.8,50.4) .. controls (193.8,51.56) and (192.86,52.5) .. (191.7,52.5) .. controls (190.54,52.5) and (189.6,51.56) .. (189.6,50.4) -- cycle ;
\draw  [color={rgb, 255:red, 0; green, 0; blue, 0 }  ,draw opacity=1 ] (196.4,50.4) .. controls (196.4,49.24) and (197.34,48.3) .. (198.5,48.3) .. controls (199.66,48.3) and (200.6,49.24) .. (200.6,50.4) .. controls (200.6,51.56) and (199.66,52.5) .. (198.5,52.5) .. controls (197.34,52.5) and (196.4,51.56) .. (196.4,50.4) -- cycle ;
\draw  [color={rgb, 255:red, 0; green, 0; blue, 0 }  ,draw opacity=1 ] (190,60.4) .. controls (190,59.24) and (190.94,58.3) .. (192.1,58.3) .. controls (193.26,58.3) and (194.2,59.24) .. (194.2,60.4) .. controls (194.2,61.56) and (193.26,62.5) .. (192.1,62.5) .. controls (190.94,62.5) and (190,61.56) .. (190,60.4) -- cycle ;
\draw  [color={rgb, 255:red, 0; green, 0; blue, 0 }  ,draw opacity=1 ] (196.8,60.4) .. controls (196.8,59.24) and (197.74,58.3) .. (198.9,58.3) .. controls (200.06,58.3) and (201,59.24) .. (201,60.4) .. controls (201,61.56) and (200.06,62.5) .. (198.9,62.5) .. controls (197.74,62.5) and (196.8,61.56) .. (196.8,60.4) -- cycle ;
\draw  [color={rgb, 255:red, 0; green, 0; blue, 0 }  ,draw opacity=1 ] (190,70.4) .. controls (190,69.24) and (190.94,68.3) .. (192.1,68.3) .. controls (193.26,68.3) and (194.2,69.24) .. (194.2,70.4) .. controls (194.2,71.56) and (193.26,72.5) .. (192.1,72.5) .. controls (190.94,72.5) and (190,71.56) .. (190,70.4) -- cycle ;
\draw  [color={rgb, 255:red, 0; green, 0; blue, 0 }  ,draw opacity=1 ] (196.8,70.4) .. controls (196.8,69.24) and (197.74,68.3) .. (198.9,68.3) .. controls (200.06,68.3) and (201,69.24) .. (201,70.4) .. controls (201,71.56) and (200.06,72.5) .. (198.9,72.5) .. controls (197.74,72.5) and (196.8,71.56) .. (196.8,70.4) -- cycle ;
\draw    (350.8,101) -- (350.8,260.01) ;
\draw    (201.8,101) -- (201.8,260.01) ;
\draw    (202.8,157.3) -- (348,160.94) ;
\draw [shift={(351,161.01)}, rotate = 181.43] [fill={rgb, 255:red, 0; green, 0; blue, 0 }  ][line width=0.08]  [draw opacity=0] (8.93,-4.29) -- (0,0) -- (8.93,4.29) -- cycle    ;
\draw    (205,206.87) -- (350,200.01) ;
\draw [shift={(202,207.01)}, rotate = 357.29] [fill={rgb, 255:red, 0; green, 0; blue, 0 }  ][line width=0.08]  [draw opacity=0] (8.93,-4.29) -- (0,0) -- (8.93,4.29) -- cycle    ;
\draw   (197,160.01) .. controls (192.33,159.9) and (189.94,162.17) .. (189.83,166.84) -- (189.74,170.58) .. controls (189.58,177.24) and (187.17,180.51) .. (182.5,180.4) .. controls (187.17,180.51) and (189.42,183.9) .. (189.25,190.57)(189.32,187.57) -- (189.17,193.84) .. controls (189.06,198.51) and (191.33,200.9) .. (196,201.01) ;
\draw    (202.6,229.7) -- (348,233.93) ;
\draw [shift={(351,234.01)}, rotate = 181.66] [fill={rgb, 255:red, 0; green, 0; blue, 0 }  ][line width=0.08]  [draw opacity=0] (8.93,-4.29) -- (0,0) -- (8.93,4.29) -- cycle    ;
\draw    (351,241.01) -- (487.01,252.75) ;
\draw [shift={(490,253.01)}, rotate = 184.93] [fill={rgb, 255:red, 0; green, 0; blue, 0 }  ][line width=0.08]  [draw opacity=0] (8.93,-4.29) -- (0,0) -- (8.93,4.29) -- cycle    ;

\draw (398.1,32) node [anchor=north west][inner sep=0.75pt]  [font=\small,xscale=0.9,yscale=0.9] [align=left] {\begin{minipage}[lt]{60pt}\setlength\topsep{0pt}
\begin{center}
wireless MIMO \\channel
\end{center}

\end{minipage}};
\draw (253,24) node [anchor=north west][inner sep=0.75pt]  [font=\small,xscale=0.9,yscale=0.9] [align=left] {\begin{minipage}[lt]{34pt}\setlength\topsep{0pt}
\begin{center}
wired \\fronthaul
\end{center}

\end{minipage}};
\draw (466.1,78.8) node [anchor=north west][inner sep=0.75pt]  [font=\small,xscale=0.9,yscale=0.9] [align=left] {User $\displaystyle k$};
\draw (298.1,77.1) node [anchor=north west][inner sep=0.75pt]  [font=\small,xscale=0.9,yscale=0.9] [align=left] {MEC servers @APs$\displaystyle \ $};
\draw (158.9,79.7) node [anchor=north west][inner sep=0.75pt]  [font=\small,xscale=0.9,yscale=0.9] [align=left] {MEC server @CPU};
\draw (355.78,122.27) node [anchor=north west][inner sep=0.75pt]  [font=\footnotesize,rotate=-353.1,xscale=0.9,yscale=0.9] [align=left] {\begin{minipage}[lt]{92pt}\setlength\topsep{0pt}
\begin{center}
1: offloading request + data
\end{center}

\end{minipage}};
\draw (211.91,135.66) node [anchor=north west][inner sep=0.75pt]  [font=\footnotesize,rotate=-353.9,xscale=0.9,yscale=0.9] [align=left] {\begin{minipage}[lt]{95pt}\setlength\topsep{0pt}
\begin{center}
2: request forwarding + data
\end{center}

\end{minipage}};
\draw (139.2,158.5) node [anchor=north west][inner sep=0.75pt]  [font=\small,rotate=-270,xscale=0.9,yscale=0.9] [align=left] {\begin{minipage}[lt]{46pt}\setlength\topsep{0pt}
\begin{center}
transmission \\+ fronthaul \\latency
\end{center}

\end{minipage}};
\draw (213.78,160.96) node [anchor=north west][inner sep=0.75pt]  [font=\footnotesize,rotate=-1.69,xscale=0.9,yscale=0.9] [align=left] {\begin{minipage}[lt]{105pt}\setlength\topsep{0pt}
\begin{center}
3: computational request + data
\end{center}

\end{minipage}};
\draw (226.02,211.84) node [anchor=north west][inner sep=0.75pt]  [font=\footnotesize,rotate=-358.35,xscale=0.9,yscale=0.9] [align=left] {\begin{minipage}[lt]{78.91pt}\setlength\topsep{0pt}
\begin{center}
4: computational output 
\end{center}

\end{minipage}};
\draw (180.4,163.23) node [anchor=south east] [inner sep=0.75pt]  [font=\footnotesize,rotate=-270,xscale=0.9,yscale=0.9] [align=left] {\begin{minipage}[lt]{49.15pt}\setlength\topsep{0pt}
\begin{center}
computational \\latency
\end{center}

\end{minipage}};
\draw (222.24,237.22) node [anchor=north west][inner sep=0.75pt]  [font=\footnotesize,rotate=-0.99,xscale=0.9,yscale=0.9] [align=left] {\begin{minipage}[lt]{78.91pt}\setlength\topsep{0pt}
\begin{center}
5: computational output 
\end{center}

\end{minipage}};
\draw (363.82,224.08) node [anchor=north west][inner sep=0.75pt]  [font=\footnotesize,rotate=-5.18,xscale=0.9,yscale=0.9] [align=left] {\begin{minipage}[lt]{78.91pt}\setlength\topsep{0pt}
\begin{center}
6: computational output 
\end{center}

\end{minipage}};

\end{tikzpicture}